\documentclass[aps,prl,twocolumn,superscriptaddress,nofootinbib, nobibnotes,floatfix,preprintnumbers,amsmath,amssymb]{revtex4-2}
\allowdisplaybreaks[1]
\usepackage{graphicx,bm,xcolor,orcidlink}
\usepackage{hyperref}
\hypersetup{colorlinks=true,citecolor=blue,linkcolor=blue,urlcolor=blue}
\newcommand{\Tr}{\mathrm{Tr}}
\newcommand{\sutwo}{$SU(2)$}
\newcommand{\NCS}{N_{\rm CS}}
\providecommand{\ket}[1]{|#1\rangle}
\newcommand{\SM}[1]{Appendix~\ref{#1}}
\newcommand{\EM}[2]{\hyperref[#1]{End Matter~#2}}
\begin{document}

\preprint{%
  \begin{minipage}[c]{0.5\textwidth}
    \includegraphics[width=0.22\textwidth]{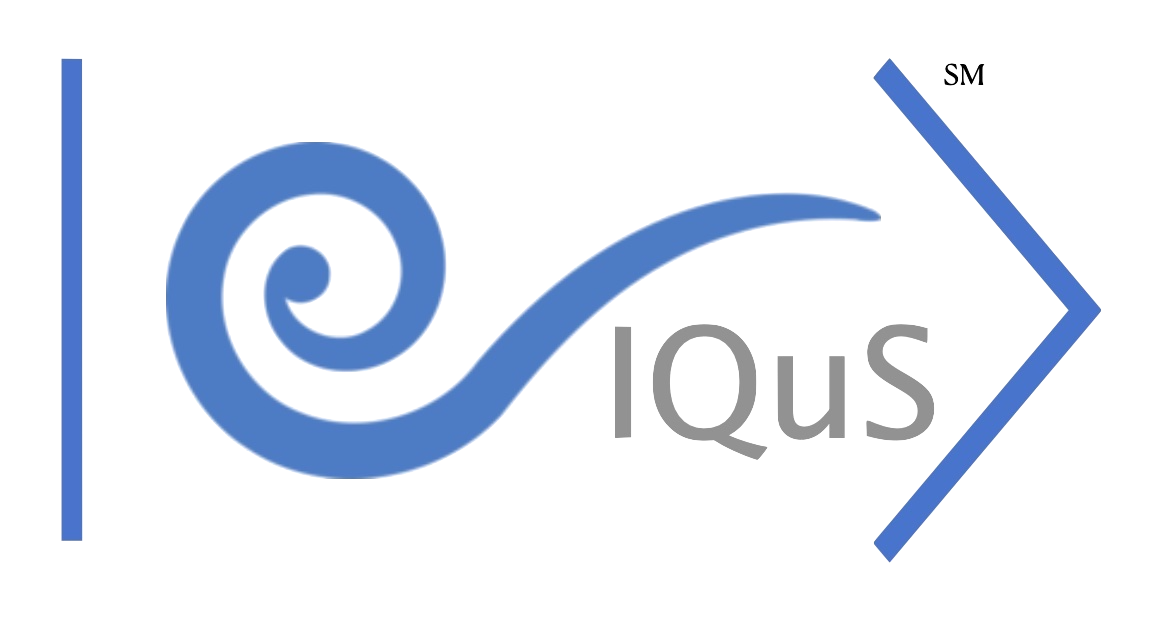}
  \end{minipage}%
  \hfill
  \begin{minipage}[c]{0.5\textwidth}
    \raggedleft IQuS@UW-21-139
  \end{minipage}}

\title{\rule{0pt}{0.6cm} Lattice Hamiltonians for Chiral Gauge Theories}

\author{Dorota M. Grabowska\,\orcidlink{0000-0002-0760-4734}}
\email{grabow@uw.edu}
\affiliation{InQubator for Quantum Simulation (IQuS), Department of Physics, University of Washington, Seattle, WA 98195}

\date{\today}

\begin{abstract}
We present a lattice Hamiltonian for chiral gauge theories on the boundary of a slab.
The formulation is gauge-fixed, so the gauge variables are the physical modes of the boundary field, and Gauss's law holds exactly.
The bulk gauge field is determined by the boundary field and an integer label, which is the topological charge that the boundary generates and the bulk stores as a localized instanton.
The bulk enters as a dielectric, and both gauge anomaly conditions become the question of whether the boundary satisfies Gauss's law on its own.
For background gauge fields, we show that sectors that differ by a unit of bulk charge are degenerate after Pauli-Villars subtraction and the mirror fermions decouple. The boundary theory has no vacuum angle.
Every term in the Hamiltonian is defined in terms of the boundary gauge variables, the integer label and the fermion modes, a starting point for quantum circuit construction.
\end{abstract}

\maketitle

\section{Introduction}
\label{sec:intro}
A non-perturbative lattice definition of chiral gauge theories is a long-standing problem~\cite{Kaplan:2009yg,Luscher:2000hn}. Its absence blocks first-principle studies of the full Standard Model and of strongly coupled chiral gauge theories.
Much recent work attempts to realize the chiral theory on the boundary of an extra dimension~\cite{Grabowska:2015qpk,Grabowska:2016bis,Kaplan:2023pxd,Kaplan:2023pvd,Kaplan:2024ezz,Sen:2026hfw}, following the domain wall construction of chiral fermions~\cite{Kaplan:1992bt}.
In these constructions both boundary and bulk fermions are dynamical.
The gauge field is dynamical on the boundary alone, and the bulk gauge field is determined by it.
Earlier constructions extend the boundary gauge field into the bulk by various prescriptions: constant in the extra dimension~\cite{Kaplan:1992bt,Golterman:1994at}, via gradient flow~\cite{Grabowska:2015qpk,Grabowska:2016bis}, or by the bulk's equations of motion~\cite{Kaplan:2024ezz}. The last two approaches were recently implemented on a Euclidean lattice~\cite{Dang:2026slab,Dang:2026disk}. As these implementations use a path integral formulation, both gradient flow and the equations of motion give a bulk at one time that depends on the boundary field at other times. The boundary's effective action is then non-local in time, and there is no transfer matrix. A Hamiltonian must behave differently, as every operator must be a function of the variables at one instant. The bulk is therefore determined by two prescriptions: how a boundary configuration is extended into the bulk, which gives the bulk links, and how the bulk's charges enter the boundary's Gauss law, which is the extension of the gauge field's time component~\cite{Sen:2026hfw}. The bulk has no dynamics of its own. It changes in time only because the boundary field and the integer label do.

In this Letter, we construct a lattice Hamiltonian that realizes chiral gauge theories, Abelian and non-Abelian, on the boundary of an extra-dimensional bulk. 
The formulation is gauge-fixed, so Gauss's law is satisfied \textit{a priori} rather than imposed \textit{a posteriori}, and holds exactly.
The bulk enters as a dielectric with a localized instanton that binds the charge expelled by the boundary.
Anomaly conditions are statements about whether the boundary can satisfy Gauss's law without help from the bulk.
Every term of the Hamiltonian is a definite function of the gauge variables, the integer label and the fermion modes, a starting point for a circuit construction built on top of the pure-gauge primitives of Ref.~\cite{Froland:2025cgt}.

Such a construction must have three properties to realize the Standard Model on the boundary.
\begin{enumerate}
\setlength{\itemsep}{0pt}
\item \textit{Boundary Weyl modes.} Any mirror partners must decouple.
A slab~\cite{Shamir:1993zy,Furman:1994ky} has a second wall that carries them, a ball has none~\cite{Kaplan:2023pxd,Kaplan:2023pvd,Clancy:2024bjb,Aoki:2024bwx,Sen:2026hfw}. We work on a slab, where the gauge fixing is simplest.
\item \textit{Bulk topological charge.} A vector theory realized this way has an exact $U(1)_A$ that anomaly inflow breaks only on the boundary~\cite{Callan:1984sa,Hamada:2017tny,Golterman:2024wxa}. The $\eta'$ is unnaturally light unless the bulk carries the topological charge out of the boundary's reach~\cite{Golterman:2024wxa,Kaplan:2024ezz}.
Sectors that differ only in bulk charge must be degenerate in energy, or else $B+L$ violation is suppressed.
\item \textit{Consistency of fermion content.} Anomalous and anomaly-free content must behave differently.
\end{enumerate}
Other candidate non-perturbative definitions of chiral gauge theories exist, which do not use an extra dimension~\cite{Luscher:1999un,Craig:2026xyz,Hsu:2026sm,Witten:2026note,Singh:2025gwh,Lamm:2026qsp,Thorngren:2026ydw,Lootens:2026cgt,Golterman:2026smg,Seifnashri:2026cgt}.

\section{The physical picture}
\label{sec:picture}
Any slab with massive bulk fermions can be interpreted as a capacitor.
Its plates are the two walls, one we call the boundary and the other we call the far wall. Weyl modes appear on both plates, and those on the far wall we call mirror modes.
The dielectric is the bulk. It screens the electric field of a boundary charge at every finite wavelength, but not the net flux through the slab, which Gauss's law ties to the total charge. Through this net flux the mirror feels the boundary's total charge and its Polyakov loops, and the boundary feels the mirror only through the Coulomb energy of its net color and the mirror's potential for the Polyakov loops. Both vanish as the volume grows.

What is particular to this construction is where the level expelled from the boundary ends up. Suppose the boundary field turns on and off again. Each flavor of boundary Weyl modes has one chirality, so the winding shifts their spectrum by one level instead of permuting it. One level per fermion flavor crosses zero and moves off the boundary into the dielectric. After turning off, the boundary gauge field is back where it began, with no record of its winding. The bulk, however, retains the history. A unit of topological charge develops within the dielectric and binds the fermion level that left the boundary. Neither the topological charge nor the fermion level reaches the mirror (Fig.~\ref{fig:geom}b).
The ground state fills that level with a fermion from the boundary's Fermi surface, so one fermion per flavor has moved into the dielectric. This is anomaly inflow.
If the bulk held no topological charge, the level would reach the mirror, as it does in the gradient-flow construction~\cite{Grabowska:2015qpk,Okumura:2016dsr}.
That it stops at the instanton instead is a proposed way past the objection of Ref.~\cite{Okumura:2016dsr}. Resolving the objection requires dynamical gauge fields and a continuum extrapolation, which this Letter does not attempt.

Whether the boundary is a gauge theory on its own is a question about Gauss's law. The whole slab satisfies it by construction. For a single flavor the fermion that moves into the dielectric takes its color with it, and the boundary's fermions are left with a net color that nothing on the boundary can cancel. The same is true for any anomalous fermion representation. For an anomaly-free representation the colors that leave form a singlet, the boundary is left a singlet, and only then does it satisfy its own Gauss's law.

In the language of condensed matter, domain wall fermions are a topological insulator.
The slab is the lattice model of the $(4+1)$-dimensional Chern insulator~\cite{Qi:2008ew}, two dimensions above the quantum anomalous Hall insulator with its chiral edge. The bulk current that carries the pumped charge to the instanton is the generalized Hall current of Ref.~\cite{Kaplan:2021ghc}.
In the Hall insulator, threading a flux quantum through the edge pumps one unit of charge off it~\cite{Laughlin:1981jd}, and the bulk's Chern-Simons response binds that charge to the flux tube.
Here the analogue of the flux is the boundary's Chern-Simons number, the bulk's response is the second Chern-Simons term, and its charge density is the instanton density, which is why the pumped charge binds to the instanton.
What the construction adds is a dynamical gauge field on the boundary. Its lower-dimensional analogue would be a Chern insulator in $2+1$ dimensions whose edge carries a dynamical gauge field.

\section{The Hamiltonian}
\label{sec:hamiltonian}
\begin{figure*}[t]
\includegraphics[width=\textwidth]{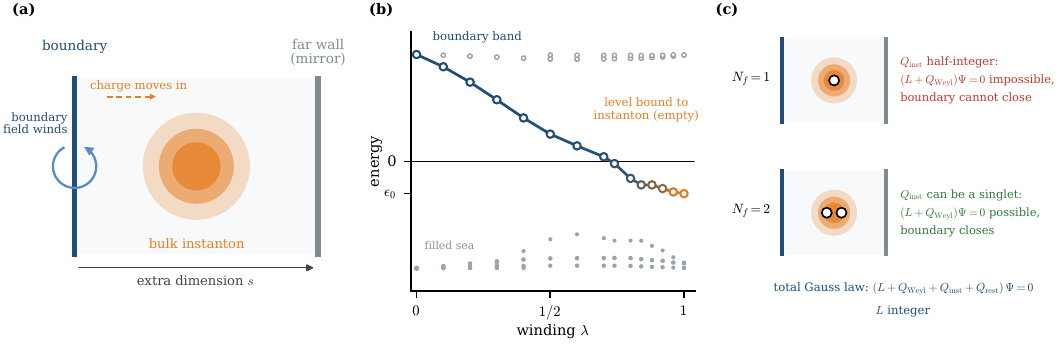}
\caption{(a)~The slab: the boundary at $s=0$ with the dynamical gauge field and the Weyl modes, the bulk instanton at mid-slab, and the flat far wall with the mirror modes. (b)~The level nearest zero along a winding, parametrized by $\lambda\in[0,1]$, on $6^3\times12$ at $M=1$, with the bulk continued along the winding and relaxed at each $\lambda$ (\SM{sm:bulkcomp}). The color of the curve is the fraction of the level's weight on the boundary layer, and the grey points are the twelve levels nearest zero, filled below zero and open above in the Dirac vacuum at each $\lambda$. The level leaves the boundary band, crosses zero, and becomes the level bound to the instanton, which the state at fixed fermion number on this lattice leaves empty, with weight below $10^{-4}$ on the far wall throughout. One curve stands for one level per flavor. (c)~Schematic of Gauss's law after a winding: the total law, Eq.~\eqref{eq:global}, always holds, and the boundary's own law, Eq.~\eqref{eq:bndyGauss}, has a low-energy solution for even $N_f$ and none for odd $N_f$.}
\label{fig:geom}
\end{figure*}

The boundary is a three-dimensional periodic lattice with links in a compact gauge group $G$, which we take to be $SU(N)$\footnote{Compact $U(1)$ has the same construction, and it is used to test the perturbative anomaly condition in \EM{em:results}{E}.}.
It is the wall at $s=0$ of the slab of Fig.~\ref{fig:geom}.
The fermion is the domain-wall fermion of Shamir's slab~\cite{Shamir:1993zy}, a Wilson fermion on an interval in $s$ with Weyl modes on the boundary and mirror Weyl modes on the far wall. We fix the gauge on a maximal tree of the boundary~\cite{Creutz:1976ch,Ligterink:2000ug}.
Tree links are set to the identity.
Each remaining link is parallel transported along the tree to the origin and is a physical variable, the rod $X_\kappa$.
A rod is invariant under every gauge transformation that is trivial at the origin, the large ones included, so the only one that survives the gauge fixing is a global $G$, acting as $X_\kappa\to hX_\kappa h^{-1}$.
The three rods that close the cycles through the origin are the Polyakov loops based there.
The links along $s$ carry no electric field, since only the boundary field is dynamical, and thus are set to the identity. This means that a bulk fermion at $(x,s)$ transforms the same way as a boundary fermion at $x$ under gauge rotations. Its charge enters Gauss's law at $x$ as well.
The bulk links parallel to the boundary are not dynamical but functions of the rods and of the integer label $n$, encoded in the map $F^{(n)}(X)$ that is developed in the next section.

The Hamiltonian, without Pauli-Villars subtraction, is
\begin{align}
H&=\frac{g^2}{2}\sum_\ell \big(\mathcal J_\ell-(D_X\Phi)_\ell\big)^2\nonumber \\
&+\frac{1}{g^2}\sum_\square \Big(N-\mathrm{Re}\Tr P_\square[X]\Big) \nonumber \\
&+\sum_{\ell}\Big(\chi^\dagger_{x}\,\Gamma_\ell\,U_\ell[X]\,\chi_{y}+{\rm h.c.}\Big) \nonumber \\
&+\sum_x\chi^\dagger_x\gamma_0(M-4r)\chi_x ,
\label{eq:H}
\end{align}
with $P_\square$ the plaquette, $\chi$ the Wilson fermion on the slab transported to the origin along the tree, $\Gamma_\ell=\tfrac12\gamma_0(\gamma_d+r)$ for $\ell$ in direction $d$, $r$ the Wilson parameter and $M$ the uniform domain-wall mass. The gauge link $U_\ell$ is 
\begin{align}
U_\ell[X]&=\begin{cases}X_\kappa&\ell\ \text{the link of rod }\kappa\\ 1&\ell\ \text{a tree link or along }s\\ F^{(n)}_\ell(X)&\ell\ \text{a bulk link}\end{cases} \, .
\label{eq:U}
\end{align}
See \SM{app:tree} for further details about maximal tree gauge fixing with fermions.

The electric term sums over the boundary links and the hopping terms sum over every link of the slab. In the gauge-fixed variables the electric flux $\mathcal J_\ell$ is a composite operator, the sum of the rod generators whose loops pass through link $\ell$ and the color charges that Gauss's law routes through it. That is how the fermions enter the electric term (\EM{em:coupling}{A}).
The shift $D_X\Phi$ is the covariant gradient of a potential $\Phi$ built from the bulk charges. The potential $\Phi$ couples the charge $Q_s$ of the bulk fermions on layer $s$ to the boundary's electric field through the heat kernel $f_s(\Delta_X)=e^{-s\Delta_X}$, with $\Delta_X$ the covariant Laplacian on the boundary~\cite{Grabowska:2015qpk,Grabowska:2016bis}.
A charge at depth $s$ then acts on the boundary as $f_s(\Delta_X)Q_s$, in full at zero transverse momentum and damped at nonzero ones, over a transverse range of $\sqrt{2s}$ lattice spacings. This is the dielectric of the physical picture.
Eq.~\eqref{eq:H} must be supplemented with Pauli-Villars terms that regulate the bulk fermions. They are added to the Hamiltonian as three functionals of the rods, a potential, a Berry connection and an inertia correction, Eq.~\eqref{eq:Hreg} of \EM{em:pv}{C}.

The gauge fixing leaves one constraint, the generator of the global gauge transformation. Only states that satisfy
\begin{align}
\big(L_{\rm tot}+Q_{\rm tot}\big)\Psi=0 ,
\label{eq:global}
\end{align}
are physical. Here $L_{\rm tot}$ is the generator of the global $G$ on the rods and $Q_{\rm tot}$ its generator on the fermions. $L_{\rm tot}$ carries no center charge, since the rods transform by conjugation.
The Hilbert space is the product of $L^2(G)$ for each rod, the Fock space of the slab fermions, and a register, a single integer $n$ that labels the bulk's topological sector. The copies of rod space at fixed $n$ are the sheets. On sheet $n$ the Hamiltonian is evaluated with the bulk links $F^{(n)}(X)$, the same $n$ that labels the wavefunction $\psi_n(X)$. The register is tied to the rods by the prescription that it change by one when the rods cross a particular surface. This allows the pair $(X,n)$ to move continuously. That surface is defined in the next section.

\section{The bulk gauge field}
\label{sec:bulk}
The bulk is determined by the boundary $X$ and an integer label $n$.
The configuration $F^{(n)}(X)$ is the stationary bulk of lowest action that holds $n$ units of topological charge, with the boundary given by $X$ and the far wall fixed by a prescription from the boundary's Polyakov loops (see \EM{em:wall}{D}). For the trivial boundary field, $X=1$, the unit of charge is a lump whose depth is a nearly flat direction, and the definition places it at mid-slab. A lump placed at mid-slab stays there for every random boundary tested (\SM{sm:bulkcomp}). Along a winding the growing topological charge is the boundary field's own excursion into the slab, held against the boundary. Only when the boundary field is off does it become a lump, which the bulk prescription then moves to mid-slab, which causes a small change of the subtracted energy (\EM{em:results}{E}).
We find each configuration by seeding the bulk with a given topological charge $n$ and relaxing with gradient flow of a bulk action $S$ in a fictitious time $\tau$,
\begin{align}
\partial_\tau U_\ell=-\partial_{U_\ell}S\,U_\ell .
\label{eq:flow}
\end{align}
The fixed point extends the boundary by the bulk's equations of motion as in Refs.~\cite{Sen:2026hfw,Dang:2026slab,Dang:2026disk}. The gradient-flow construction of Refs.~\cite{Grabowska:2015qpk,Grabowska:2016bis} instead uses $s$ itself as the flow time.
The flow must hold a lump of unit charge stationary.
Which action does so, at what lump size and on what lattice, is a question of implementation (\SM{app:numerics}).
In the numerical tests of this Letter the lumps are a few spacings across and are held by L\"uscher's action~\cite{Luscher:1998du}.

The stationary field obeys an elliptic equation of second order in $s$.
A boundary mode of transverse momentum $k$ decays into the bulk as $e^{-ks}$ at linear order, and at a rate set by the covariant Laplacian beyond it.
The Polyakov loops, the zero modes of that Laplacian about a flat boundary, extend as constants, and they are all the mirror sees of the boundary.
The prescription for the far wall is continuous except on a set of codimension two, which leaves no mark on the fermion sea in the loops computed with an interpolated bulk (\SM{app:wall}).

Topological charge is a four-dimensional quantity, and a three-dimensional field has only a Chern-Simons number. A path of boundary fields $X(\lambda)$, with $\lambda$ as a fourth coordinate, is a four-dimensional configuration whose charge is the change of $\NCS$ along it. A winding, the boundary field turning on and off again as in Sec.~\ref{sec:picture}, carries one unit,
\begin{align}
\int_0^1 d\lambda\sum_x q(x,\lambda)=1 ,
\label{eq:winding}
\end{align}
with $q$ L\"uscher's geometric topological density~\cite{Luscher:1981zq} and $\NCS[X]$ the partial integral up to $X$, defined mod 1 (\EM{em:register}{B}).
At the end of a winding the boundary field is back where it began, so the unit has to sit in the bulk, as a localized instanton. A winding returns the boundary to $X=1$ and leaves the bulk at a second stationary point, one that holds a unit of topological charge. The bulk seeded from that charge alone reaches the same point, up to the translation to mid-slab. A winding followed by its inverse leaves no trace (\EM{em:results}{E}).
The bulk therefore remembers one thing about the boundary's history, the number of windings $n$, and $F^{(n)}(X)$ depends on the boundary through $X$ and $n$ alone.
The slab's topological charge is $n$ from the instanton plus $\NCS[X]$ from the boundary, with $\NCS$ on the branch $(-\tfrac12,\tfrac12]$.
The register changes by one where $\NCS$ crosses the edge of that branch, the sphaleron surface, so the slab's charge is continuous and the instanton is the only record of a winding (\SM{sm:register}).

The rods range over all of their configuration space, with no admissibility condition. Short loops around the set on which $\NCS$ mod 1 is undefined change $n$ as a winding does, at a magnetic cost of order $1/g^2$, and their rate is not computed here (\SM{sm:register}).
A lattice-scale plaquette dislocation jumps the integral of Eq.~\eqref{eq:NCS} by an integer. That jump is not a crossing of the sphaleron surface and the register does not count it, so $\NCS$ on its branch, $n$ and the bulk are all unchanged (\SM{sm:register}).
\begin{align}
F^{(n)}(hXh^{-1}) = h\,F^{(n)}(X)\,h^{-1}, \qquad h\in G, \label{eq:equivariant}
\end{align}
without which the generator of the total Gauss law, Eq.~\eqref{eq:global}, would not commute with the Hamiltonian.
The construction respects the one symmetry gauge fixing leaves. Rotating all the rods by $h$ turns the instanton's color orientation by the same $h$, which is what Eq.~\eqref{eq:global} and the argument of Sec.~\ref{sec:anomalies} require, except on sets of codimension one, which a rod path crosses and on which two orientations tie and the choice jumps (\SM{sm:orientation}).

\paragraph{\textbf{Decoupling and degeneracy of the sheets}}
Along a winding one level per flavor crosses zero and becomes the level the instanton binds, at energy $\epsilon_0$ inside the bulk fermion's mass gap (Fig.~\ref{fig:geom}b). It is localized on the instanton rather than paired with the boundary Weyl fermion, and reaches the boundary only through the Polyakov loops. On the $L_b^3\times L_s$ slabs computed, with $L_s=12$ and $16$, its weight on the far wall is below $10^{-4}$ at every point of the winding, falling by nearly a factor of ten per layer (\EM{em:results}{E}).
The instanton raises the energy of the filled sea by an amount of order the bulk mass.
That energy belongs to the bulk vacuum, and the Pauli-Villars partners that regulate the bulk fermions remove it. The partners live on a closed lattice, the slab glued to its reflection~\cite{Kikukawa:2001mw}, so that their operator has no boundary (\EM{em:pv}{C}).
A field of the wrong statistics has no positive-norm Hamiltonian, so the partners enter as functions of the rods, a potential, a Berry connection and a correction to the rods' inertia. These expressions, Eq.~\eqref{eq:Hreg}, are exact for a static boundary field (\EM{em:pv}{C}, \SM{app:pv}).

These three terms are not the whole story once the rods are dynamical. With the partners alone integrated out, the rods' kinetic term in Eq.~\eqref{eq:Hreg} stays positive only for $g^2L_s$ of order a few on the volumes computed, and above that the Hamiltonian is unbounded from below. The deep bulk of the fermion sea responds to the same modes with the opposite sign. With both seas integrated out together the kinetic term stays positive on the volumes computed. Linearized about $X=1$ it is positive and independent of $L_s$ on the volumes computed, up to a coupling above the scaling window. With the layer closure, the reflection taken about the first and last layers instead of across the links beyond them, it is positive for every $g^2$ (\EM{em:pv}{C}). A functional of both seas with this property is the regulated Hamiltonian for dynamical rods, and its construction is left for future work. Every static result below is independent of this, since the inertia enters only for dynamical rods. 

After subtraction the Dirac vacua of the two sheets are degenerate, to $3\times10^{-4}$ at $M=1$ and to $2\times10^{-3}$ across the range of $M$ computed. The energy barrier along the winding computed is the magnetic energy of the boundary field at its midpoint, $13.6/g^2$ at the winding's radius $2.5$, independent of $L_b$ from $6^3$ on and an upper bound on the sphaleron energy. The subtracted sea adds $0.080$ to it (\EM{em:results}{E}).
The wound vacuum has one more fermion than the trivial one, so the state a winding reaches at fixed fermion number has a hole.
In a large volume the hole sits at the Fermi surface of the boundary's gapless Weyl sea with the bound level filled, so the sheets are degenerate at fixed fermion number up to $1/L_b$. On the small lattices computed the boundary has no level below $|\epsilon_0|$, so the hole sits on the bound level and the splitting is its energy (\EM{em:results}{E}). No symmetry of the Hamiltonian relates the sheets.

\paragraph{\textbf{The label and the vacuum angle}}
Without a bulk, the only record of a winding is a phase $e^{i\theta}$ around the closed path, the $\theta$ vacuum~\cite{Callan:1976je,Jackiw:1976pf}.
Here a winding is not a closed path.
It takes the bulk from $F^{(n)}$ to $F^{(n+1)}$ and moves one boundary fermion per flavor onto the bound level.
A phase $e^{i\theta}$ could enter only in the matching of the wavefunction across the sphaleron surface, and the unitary $e^{i\theta\hat n}$, with $\hat n$ the register, removes it.
Since a winding changes the boundary's fermion number, that unitary acts on the boundary as its anomalous $U(1)$ rotation (\SM{sm:register}).
The boundary theory therefore has no vacuum angle of its own, for any content with a Weyl fermion.
Whether a vector-like theory realized this way also loses the phase of its fermion mass, as proposed in Ref.~\cite{Kaplan:2024ezz} and disputed in Ref.~\cite{Golterman:2024wxa}, is the $\eta'$ question of the Outlook. A simulation of this Hamiltonian could answer this question.

\section{Anomaly conditions as Gauss's law}
\label{sec:anomalies}
Eq.~\eqref{eq:global} has solutions for any fermion content, since the slab as a whole is a gauge theory by construction.
The question is whether the boundary is one on its own.
Splitting Eq.~\eqref{eq:global} into four contributions,
\begin{align}
\big(L_{\rm tot}+Q_{\rm Weyl}+Q_{\rm inst}+Q_{\rm rest}\big)\Psi=0 
\label{eq:split}
\end{align}
separates the sources of color: $L_{\rm tot}$ is the color of the rods, $Q_{\rm Weyl}$ that of the light modes localized on the boundary, $Q_{\rm inst}$ that of the level bound to the instanton, and $Q_{\rm rest}$ that of the massive modes and the mirror. For \sutwo{} $L_{\rm tot}$ is integer.

The chiral gauge theory is the theory of the rods and the Weyl fermions alone.
It exists as a gauge theory when its own Gauss's law,
\begin{align}
(L_{\rm tot}+Q_{\rm Weyl})\Psi=0 ,
\label{eq:bndyGauss}
\end{align}
has a solution among the low-energy states of every sector its dynamics reach, so that the boundary is never carried out of its singlet sector. A state is defined to be in the low-energy spectrum when its energy above the trivial sheet vanishes as $L_b\to\infty$.
By Eq.~\eqref{eq:split} this is the condition that $Q_{\rm inst}+Q_{\rm rest}$ be a singlet in a low-energy state, at every instant along any path of boundary configurations.
States in which $Q_{\rm inst}+Q_{\rm rest}$ carry color and the boundary carries the opposite color exist whenever bound levels are occupied, split from the singlet by the Coulomb energy of that color.
The two anomaly conditions are the two ways the singlet state can fail to exist.

The perturbative condition is local.
The massive modes have a filled Dirac sea, and under a change of the boundary field that sea responds with a gauge charge on the layers nearest the boundary, a part of $Q_{\rm rest}$, with the same coefficient as the corresponding triangle anomaly for that fermion representation.
This is the charge of anomaly inflow, the massive sea's response near the boundary, and it is the same for any extension of the boundary field into the bulk~\cite{Grabowska:2015qpk,Grabowska:2016bis}. We test the $1+1$-dimensional analogue of this condition in a $U(1)$ slab with a simplified bulk (\EM{em:results}{E}).
For anomaly-free fermion representations that charge vanishes in every background.
Nothing crosses at the far wall along a winding either, so $Q_{\rm rest}$ is then a singlet at every instant.
For an anomalous fermion representation $Q_{\rm rest}$ is not a singlet for a generic background, and Eq.~\eqref{eq:bndyGauss} fails there.
For \sutwo{} the coefficient vanishes identically, consistent with the statement that pure \sutwo{} has no triangle anomaly.

Instead, \sutwo{} has a global anomaly~\cite{Witten:1982fp} and here it appears only under a winding.
The global anomaly is usually stated as a sign picked up by the vacuum around a non-contractible loop of gauge transformations~\cite{Nelson:1984gu,Neuberger:1998xn,Bar:2000qa,Wang:2018qoy}, or as the statement that $(-1)^F$ is the center of \sutwo{}. Both imply that an odd number of fermions per instanton is inconsistent.
In this construction it is the statement that, after a winding, no low-energy state satisfies the boundary's own Gauss's law.
At fixed fermion number, with $N_f$ flavors of \sutwo{} doublets, the wound sheet has states with $k$ of the bound levels occupied and the boundary $k$ fermions short, for every $k$ from $0$ to $N_f$. In a large volume a boundary hole costs $O(1/L_b)$, so such a state lies $(N_f-k)|\epsilon_0|$ above the trivial sheet.

An occupied bound level carries the center charge of the doublet it came from, and by Eq.~\eqref{eq:equivariant} it rotates with the rods. With $k$ levels occupied the boundary's color is integer when $k$ is even. When $k$ is odd it is half-integer, which the integer rods cannot neutralize. The states that close the boundary's own Gauss's law are therefore those with $k$ even. Taking a fermion from the mirror instead puts a half-integer color at the far wall that the boundary cannot create or remove, since the mirror's fermion number changes only by hopping to the instanton, with an amplitude that falls exponentially in $L_s$. The boundary feels that charge only at zero momentum, as a constant background field whose field lines come from nowhere it can act on, so this is a sector its dynamics reach only with that amplitude, and the mirror stays in its sea.
For even $N_f$ the state with all $N_f$ levels occupied is among them, and it lies in the low-energy spectrum, degenerate with the trivial sheet in a large volume.
For odd $N_f$ the largest even $k$ is $N_f-1$, so every state that closes the boundary's Gauss's law lies at least $|\epsilon_0|$ above the trivial sheet, and no state of the wound sheet both closes it and lies in the low-energy spectrum. A winding carries the boundary out of its singlet sector, and the chiral gauge theory does not exist as a theory of $L_{\rm tot}$ and $Q_{\rm Weyl}$ alone.
The boundary feels this through the one thing the dielectric cannot screen, the zero-momentum Coulomb energy of its half-integer color $j$, $g^2 j(j+1)/(V\mu_{\rm min})$ with $V=L_b^3$ and $\mu_{\rm min}$ the lowest eigenvalue of $\Delta_X$ in the given background. That eigenvalue vanishes only at a reducible boundary and is of order $1/L_b^2$ otherwise, so the term is a potential on the Polyakov loops that is absent for even $N_f$.

\section{Outlook}
\label{sec:outlook}
A lattice Hamiltonian that realizes chiral gauge theories on a boundary has to have three properties. In this Letter we have shown that, for background gauge fields on the lattices computed, the construction presented here has all three.
The mirror modes have a mechanism to decouple. The boundary field reaches the far wall only through its Polyakov loops, and in every background computed the level that crosses zero carries exponentially small weight there.
The bulk holds the topological charge the boundary generates, as an instanton at mid-slab that boundary operators reach only through the Polyakov loops. Sheets that differ by one unit of it are degenerate once the Pauli-Villars partners are subtracted, and the barrier along a winding is the boundary field's own magnetic energy. Nothing in the bulk suppresses $B+L$ violation.
Anomalous and anomaly-free content are distinguished by Gauss's law. The boundary satisfies its own Gauss's law in the low-energy spectrum only when the content is anomaly free, and both anomaly conditions take this form.

Two remaining questions about this construction require a dynamical gauge field. The first is whether the conserved charge of Ref.~\cite{Golterman:2024wxa} breaks with a propagating Goldstone boson or its Ward identity is saturated by the $\eta'$ zero mode alone~\cite{Kaplan:2024ezz}. The second is whether a field coupled to the boundary topological density acquires a potential~\cite{Ai:2026rev,Benabou:2025clr}.
Simulating this Hamiltonian could address both.
Left for future work are implementing the same construction on the ball geometry, treating the deep bulk of the fermions and their partners as one functional, testing the independence of boundary quantities from the action that holds the instanton, and constructing circuit elements beyond the pure-gauge primitives of the mixed basis~\cite{DAndrea:2023qnr,Grabowska:2024emw,Froland:2025cgt}.

Lastly, note that gauge fixing is a choice of coordinates, not a requirement. In a Kogut-Susskind formulation every boundary link is dynamical and the physical states are those that satisfy Gauss's law with each bulk fermion counted at its boundary site. The bulk is obtained from the links by the same flow, which is gauge covariant, and the register's prescription can be stated for the gauge-invariant charge $n+N_{\rm CS}$ (\SM{app:tree}).

\begin{acknowledgments}
We gratefully acknowledge discussions with David B. Kaplan, Sebastian Grieninger, Henry Lamm and Michael Wagman.
This work was supported, in part, 
by U.S. Department of Energy, Office of Science, Office of Nuclear Physics, InQubator for Quantum Simulation (IQuS)\footnote{\url{https://iqus.uw.edu}} under Award Number DOE (NP) Award DE-SC0020970 via the program on Quantum Horizons: QIS Research and Innovation for Nuclear Science\footnote{\url{https://science.osti.gov/np/Research/Quantum-Information-Science}}.
It was also supported, in part, by the Department of Physics\footnote{\url{https://phys.washington.edu}}
and the College of Arts and Sciences\footnote{\url{https://www.artsci.washington.edu}} at the University of Washington. 
We have made extensive use of Wolfram {\tt Mathematica}~\cite{Mathematica} and {\tt python}~\cite{python3,Hunter:2007} in the {\tt Conda} environment~\cite{anaconda}.

\textit{AI-use disclosure:} Claude (Anthropic) was used throughout this work. The ideas are the author's. Claude generated much of the code from the author's earlier Mathematica codes and specifications, and produced the figures to the author's design. Claude also assisted in drafting and editing the manuscript from the author's outline and specifications. All text was heavily revised by the author, who takes full responsibility for the content.

\end{acknowledgments}

\bibliography{refs}

\section*{End Matter}

\subsection*{A. The screened electric coupling}
\label{em:coupling}
On a tree link $\ell$ the electric flux has contributions from rods whose loops cross it and the color charge beyond it along the tree,
\begin{align}
\mathcal J^a_\ell&=\sum_{\kappa\in t_+(\ell)}\mathcal E^a_{L}(\kappa)-\sum_{\kappa\in t_-(\ell)}\mathcal E^a_{R}(\kappa)-Q^a_{S(\ell)} ,
\label{eq:J}
\end{align}
where $\mathcal E_{L,R}(\kappa)$ are the left and right generators of the rod $X_\kappa$, $t_\pm(\ell)$ the rods whose loops traverse $\ell$ in either direction, and $Q_{S(\ell)}$ the color charge beyond $\ell$ along the tree.
On a rod $\kappa$ the flux is $\mathcal E_L(\kappa)$ with no charge term (\SM{sm:transportedE}).
The shift $D_X\Phi$ of Eq.~\eqref{eq:H} is
\begin{align}
\Phi=\sum_{s\ge1} w_s(\Delta_X)\,Q_s,\qquad &w_s(\mu)=\frac{1-f_s(\mu)}{\mu}, \nonumber \\
&\;f_0=1,
\label{eq:Phi}
\end{align}
with $Q_s$ the color charge of the fermions on layer $s$, $D_X$ the covariant difference on the boundary acting on adjoint fields, $(D_X\Phi)_\ell={\rm Ad}(U_\ell)\Phi_y-\Phi_x$ for $\ell$ from $x$ to $y$, and $\Delta_X=D_X^{T}D_X$ the covariant Laplacian with eigenvalues $\mu$.
At $\mu=0$, $w_s(0)=s$ is finite, and the zero modes of $\Delta_X$ drop out of $D_X\Phi$. This is why a charge at any depth is seen in full at zero transverse momentum. 
At nonzero transverse momentum the kernel reduces the coupling of a charge on layer $s$ to the boundary's electric field by $e^{-s \mu}$, so the deep bulk and the mirror couple to the boundary gauge field only through its smoothest modes. This is the form factor of Refs.~\cite{Grabowska:2015qpk,Grabowska:2016bis}, where the fermions on layer $s$ see the boundary field flowed for time $s$.
The function $f_s$ is the Gauss-law multiplier extended into the bulk, and $f_0=1$ because at $s=0$ it is the boundary's own multiplier. The boundary's Gauss law is unchanged.
The shift is a choice of how the bulk charges couple to the boundary, not a rewriting of Eq.~\eqref{eq:H}.
Any $f_s$ with $f_0=1$ gives a consistent Gauss law, and the generator is unchanged (\SM{sm:screened}).

\subsection*{B. The register and the sheets}
\label{em:register}
The Chern-Simons number of the boundary is the partial integral of Eq.~\eqref{eq:winding},
\begin{align}
\NCS[X]=\int_0^{\lambda_X} d\lambda\sum_x q(x,\lambda)\ \ \mathrm{mod}\ 1,
\label{eq:NCS}
\end{align}
along any path from the trivial boundary to $X=X(\lambda_X)$.
Here $q$ is L\"uscher's geometric density~\cite{Luscher:1981zq} of the four-dimensional configuration $(x,\lambda)$ with trivial links along $\lambda$.
L\"uscher constructs transition functions from the links, and $q$ is the Chern-Simons integral of those functions over the boundary of each hypercube.
Its total over a closed lattice is an integer for every configuration outside an exceptional set of measure zero, which is why closed paths give integers and $\NCS$ is defined mod 1. L\"uscher's bound keeps a field away from that set, so that the charge is also constant along paths.
Rod space is simply connected, so every surface that spans a loop along which $\NCS$ advances by one meets the set on which $\NCS$ mod 1 is undefined, and the sheets are the covering space of rod space with that set removed. We assume that set has codimension at least two in rod space. This is not proven here.
The path of a winding is closed in rod space because every Wilson loop of a pure gauge field is one.
The instanton's depth is a nearly flat direction of the action, and near either wall the instanton binds a wall state, which is why the definition of $F^{(n)}$ fixes its depth at mid-slab (\SM{sm:register}, \SM{sm:bulkcomp}).

\subsection*{C. The Pauli-Villars terms}
\label{em:pv}
For a boundary field that varies slowly compared with the bulk gap, the partners' determinant is a derivative expansion in the rods' velocities. Its first three terms are a potential, a Berry connection and an inertia correction, and the Hamiltonian of the rods is
\begin{align}
H_{\rm reg}=H\Big|_{\mathcal E_L(\kappa)\to\mathcal E_L(\kappa)-A^{\rm PV}_\kappa,\ \ \mathcal G\to\mathcal G+\mathcal I^{\rm PV}}+V^{\rm PV}.
\label{eq:Hreg}
\end{align}
The potential $V^{\rm PV}$ is added to the Hamiltonian. The connection $A^{\rm PV}_\kappa$ shifts the generator of each rod, as a vector potential shifts a momentum. The inertia correction $\mathcal I^{\rm PV}$ adds to the inertia $\mathcal G$ of the rods. Here $\mathcal G$ is the mass matrix of the rods, whose inverse $\mathcal G^{-1}=g^2(1+T^{\rm T}T)$ is the quadratic form of the electric term of Eq.~\eqref{eq:H} in the rod generators, and $T$ writes the flux of each tree link as the sum of the rod fluxes through it. The charge terms and the shift $D_X\Phi$ of the electric term are unchanged.
$V^{\rm PV}[X]$ is minus the partner sea's ground-state energy, $A^{\rm PV}[X]$ its Berry connection on the rods, and $\mathcal I^{\rm PV}[X]$ its quantum geometric tensor, each a function of the rods through $F^{(n)}(X)$.
The connection enters only through its curvature, and the inertia correction is bounded together with the fermion sea's, while alone it grows with the slab's depth $L_s$ (\SM{app:pv}).
Eq.~\eqref{eq:Hreg} is bounded from below only where $\mathcal G+\mathcal I^{\rm PV}>0$. On the volumes computed that fails for the stiff Coulomb modes, the rod directions of largest electric energy, at $g^2L_s$ of order a few, and above it the Hamiltonian with the partners alone as a functional is unbounded from below.
For those modes the fermion sea responds with the opposite sign, and with the deep bulk of both seas as one functional the condition is $\mathcal G+\mathcal I^f+\mathcal I^{\rm PV}>0$, independent of $L_s$.
With the link closure of Appendix~\ref{sm:pvop} it holds for $g^2$ below about $30$ on $3^3$ and $11$ on $4^3$. With the layer closure the inertia of both seas is positive definite, so it holds for every $g^2$, at the price of a larger holonomy dependence of the static sea energy (\SM{sm:pvdyn}).
The construction of the functional of both seas is left for future work.
The partners live on the slab glued to its reflection~\cite{Kikukawa:2001mw}, so that their operator closes without a second boundary. The closed lattice holds the instanton and its image and binds a level at each, at $\pm|\epsilon_0|$, so the subtraction removes the bulk's response to the instanton while leaving the instanton itself.

\subsection*{D. The far-wall prescription}
\label{em:wall}
The far wall is flat, so it is fixed by its three commuting holonomies, which the construction sets from the boundary's Polyakov loops.
For \sutwo{}, let $W_i(x)=\exp(i\,\phi_i(x)\,\hat\eta_i(x)\cdot\sigma)$ be the loop around the cycle in direction $i$ through boundary site $x$, transported to the origin.
Average over the sites, $\bar b_i=\langle\sin\phi_i\,\hat\eta_i\rangle_x$ and $\bar c_i=\langle\cos\phi_i\rangle_x$, let $\hat\eta$ be the principal axis of the three vectors $\bar b_i$, the unit vector that maximizes $\sum_i(\bar b_i\cdot\hat\eta)^2$, and set
\begin{align}
W^{\rm wall}_i&=\exp\big(i\,\varphi_i\,\hat\eta\cdot\sigma\big),\nonumber\\
\varphi_i&=\mathrm{atan2}\big(\bar b_i\cdot\hat\eta,\,\bar c_i\big)\,.
\label{eq:wallks}
\end{align}
The prescription is covariant and depends on no single base point. On every computed winding it gives a wall within $0.08$ of the identity. The computations of Appendix~\ref{app:numerics} used the identity wall in its place (\SM{app:wall}).

\subsection*{E. Numerical checks}
\label{em:results}
Three properties carry the claims of this Letter: the crossing level has no weight on the far wall (decoupling), the sheets are degenerate once the Pauli-Villars subtraction is included (no bulk barrier), and exactly one level per flavor crosses zero (the color of the bound level).
All three were checked in background boundary fields for \sutwo{}, on slabs $L_b^3\times L_s$ with boundary size $L_b=6$ and depth $L_s=12$ and $16$, at $M=1$ with the bulk instanton at mid-slab and the far wall at the identity (\SM{app:numerics}).
The instantons are a few spacings across, their size set by the action of the flow, and their continuum limit is not taken here.
A sector-$n$ bulk was found on every boundary configuration studied, including random boundaries with net flux through a plane.
On such boundaries the bound level lies between $-0.30$ and $-0.45$, with $99.8\%$ of its weight on the instanton's core, layers $3$ to $8$ (\SM{sm:bulkcomp}).
Over the trivial boundary the wound bulk's action changes by one percent across the depths $3.6$ to $7.5$ of $12$ layers, and with the wall held flat a lump does not move once placed.
The crossing level ends on the instanton's core at $\epsilon_0=-0.27$, inside the bulk gap and off both walls for $0.6\lesssim M\le2$, with weight below $10^{-4}$ on the far wall at every point of the winding (Fig.~\ref{fig:geom}b).
After the Pauli-Villars subtraction the Dirac vacua of the two sheets agree to $3\times10^{-4}$ at both $L_s$, and to $2\times10^{-3}$ at $M=0.5$, $1$ and $2$ with the instanton at depth $3.6$, and the subtracted electric sheet difference is below $10^{-4}$ with any damped coupling.
Along the winding the subtracted sea energy is $0.080$ at the sphaleron, against the boundary field's magnetic energy of $13.6/g^2$ there, and $-0.004$ at the end of the path, before the final relaxation and the translation to mid-slab.
These are static energies along a path. Transition rates between the sheets need the rod dynamics and are not computed here.
A winding and its inverse, traversed together or one after the other, restore the trivial bulk and spectrum to $10^{-9}$ in the action and charge.
The bulk seeded from the boundary alone agrees with the bulk continued along the winding to $10^{-8}$ in the action, at the end of the winding and at three interior points of it (\SM{sm:bulkcomp}).
The two state types of Sec.~\ref{sec:anomalies} are both seen: with antiperiodic boundary fermions the wound state at fixed fermion number has the bound level empty and lies $|\epsilon_0|$ above the trivial sheet, and with periodic ones the hole moves to a boundary zero mode, the bound level fills, and the two sheets are degenerate (\SM{app:numerics}).
The perturbative condition was tested separately, in a $U(1)$ slab with a $1+1$-dimensional boundary and a simplified bulk. Along a winding the bulk acquires gauge charge $\lambda$ for one Weyl fermion of charge one and below $3\times10^{-4}$ for the anomaly-free $3$-$4$-$5$-$0$ content, so the boundary's own Gauss law is sensitive to how the bulk is built exactly when the anomaly does not cancel (\SM{sm:results}).

\clearpage
\onecolumngrid
\begin{center}{\large\bf Supplemental Material}\end{center}

\appendix
\setcounter{secnumdepth}{2}

\section{Fermions in the maximal tree}
\label{app:tree}

This appendix extends the maximal tree gauge fixing of Ref.~\cite{DAndrea:2023qnr} to fermions in the fundamental representation, on the boundary and in the bulk.
The maximal tree construction for pure gauge theory goes back to Refs.~\cite{Creutz:1976ch,Muller:1983rs,Bronzan:1984xb}, and quarks were included for $SU(3)$ in $2+1$ dimensions in Ref.~\cite{Bronzan:1992wc}, with a loop convention specific to that dimension.
The construction below holds in any dimension.
It gives the operators of Eqs.~\eqref{eq:J} and \eqref{eq:global}, the transformation law of the bulk field, Eq.~\eqref{eq:equivariant}, the screened coupling and the register.
The numerical tests are in Appendix~\ref{sm:maxtreenum}.

\subsection{Gauss's law with fermions}
\label{sm:gausslaw}
We start from the Kogut-Susskind Hamiltonian~\cite{Kogut:1974ag} on a $d$-dimensional lattice with sites $x$ and links $\ell(x,e_i)$, and add a fermion $\psi(x)$ in the fundamental representation on every site, with spinor and flavor indices suppressed.
The link operators satisfy
\begin{align}
[E^a_L(\ell),U(\ell)]=T^a\,U(\ell)\,,\qquad [E^a_R(\ell),U(\ell)]=U(\ell)\,T^a\,,
\end{align}
and the gauge transformation $U(x,e_i)\to\Omega(x)\,U(x,e_i)\,\Omega(x+e_i)^\dagger$ is generated at site $x$ by
\begin{align}
\label{sm:G}
G^a(x)=\sum_{i=1}^d\left[E^a_R(x-e_i,e_i)-E^a_L(x,e_i)\right] .
\end{align}
The fermion transforms as $\psi(x)\to\Omega(x)\psi(x)$, generated by the color charge
\begin{align}
\label{sm:Qpsi}
Q^a(x)=\psi^\dagger(x)\,T^a\,\psi(x)\,,\qquad
[Q^a(x),\psi(x)]=-T^a\,\psi(x)\,,\qquad [Q^a(x),\psi^\dagger(x)]=\psi^\dagger(x)\,T^a\,,
\end{align}
and couples to the links through the hopping terms
\begin{align}
\mathsf h(x,e_i)=\psi^\dagger(x)\,\Gamma_i\,U(x,e_i)\,\psi(x+e_i)\,,
\end{align}
with $\Gamma_i$ acting on spinor indices only. The on-site terms $\psi^\dagger(x)\gamma_0(M-4r)\psi(x)$ are gauge invariant on their own.

The relative sign of $G^a$ and $Q^a$ in Gauss's law follows from requiring that the full generator commute with the hopping terms.
$G^a(x)$ contains $-E^a_L(x,e_i)$ and $G^a(x+e_i)$ contains $+E^a_R(x,e_i)$, so
\begin{align}
[G^a(x),\mathsf h(x,e_i)]&=-\psi^\dagger(x)\,\Gamma_i\,T^a\,U(x,e_i)\,\psi(x+e_i)=-[Q^a(x),\mathsf h(x,e_i)]\,,\nonumber\\
[G^a(x+e_i),\mathsf h(x,e_i)]&=\psi^\dagger(x)\,\Gamma_i\,U(x,e_i)\,T^a\,\psi(x+e_i)=-[Q^a(x+e_i),\mathsf h(x,e_i)]\,,
\end{align}
and Gauss's law is
\begin{align}
\label{sm:gauss}
\left[G^a(x)+Q^a(x)\right]\ket{\Psi_\text{phys}}=0\qquad\forall\,x\,.
\end{align}

\subsection{Transported fermions}
\label{sm:transported}

The gauge fixing follows Ref.~\cite{DAndrea:2023qnr}.
Links on the maximal tree are tree links, all others are physical links, denoted $\kappa$.
For each site $x$ let $\mathcal P(x)$ be the unique path along the tree from the origin $x_0$ to $x$, and $W(x)$ the ordered product of the link operators along it.
Each physical link $\kappa(x,e_i)$ defines the loop $P(\kappa)$ from the origin along $\mathcal P(x)$, across $\kappa$, and back along $\mathcal P(x+e_i)$. The loop operator
\begin{align}
X_\kappa=W(x)\,U(x,e_i)\,W(x+e_i)^\dagger
\end{align}
is the rod of the main text, and each loop contains exactly one physical link.

The fermion is transported to the origin along the tree in the same way,
\begin{align}
\label{sm:chi}
\chi(x)=W(x)\,\psi(x)\,.
\end{align}
Since $W(x)$ is unitary, the $\chi$ are canonical fermions.
Like the rods they transform only under the gauge transformation at the origin, $\chi(x)\to\Omega(x_0)\,\chi(x)$, and their color charge is $Q^a(x)$ transported to the origin,
\begin{align}
\label{sm:WTW}
\chi^\dagger(x)\,T^a\,\chi(x)=W^1_{ab}(x)\,Q^b(x)\,,\qquad W^\dagger\,T^a\,W=W^1_{ab}\,T^b\,,
\end{align}
where the second equation defines the adjoint matrix $W^1$ of $W$, which is real and orthogonal.
On the gauge-fixed section every tree link is the identity and $\chi(x)=\psi(x)$, but the two are different operators, since the electric operators of the tree links act on $W(x)$. That action produces the charge term below.

In terms of the transported fermions the hopping terms are
\begin{align}
\mathsf h(\ell)=\chi^\dagger(x)\,\Gamma_i\,\chi(x+e_i)\quad(\ell\ \text{a tree link})\,,\qquad
\mathsf h(\kappa)=\chi^\dagger(x)\,\Gamma_i\,X_\kappa\,\chi(x+e_i)\quad(\kappa\ \text{a physical link})\,,
\end{align}
since for a tree link one of the two paths is the other extended by the link itself, so $W(x)\,U(x,e_i)\,W(x+e_i)^\dagger=1$, and for a physical link the same product is $X_\kappa$.
The hopping terms stay nearest-neighbor, with a rod on a physical link and nothing on a tree link, as in the plaquettes.

\subsection{Transported electric operators}
\label{sm:transportedE}

Let $\ell(x,e_i)$ be a tree link oriented outward from the origin, with $x$ the end nearer the origin.
Its transported electric operator is
\begin{align}
\label{sm:Jdef}
\mathcal J(\ell)=W(x)\,E_L(\ell)\,W(x)^\dagger\,,\qquad
\mathcal J^a(\ell)=W^1_{ab}(x)\,E^b_L(\ell)\,.
\end{align}
If the closer end is $x+e_i$, $\mathcal J(\ell)=-W(x+e_i)\,E_R(\ell)\,W(x+e_i)^\dagger$ takes its place and nothing below changes.
For a physical link $\kappa$ the same definition gives the left generator of the rod $X_\kappa$, which commutes with every other rod and every transported fermion. We write $\mathcal E_L(\kappa)$ for it and $\mathcal E_R(\kappa)$ for the right generator, $[\mathcal E^a_R(\kappa),X_\kappa]=X_\kappa T^a$, the $\mathcal E_{L,R}(\kappa)$ of Eq.~\eqref{eq:J}.

A tree path contains each tree link at most once, so for a fermion at $x'$ either $\ell$ is on $\mathcal P(x')$ or it is not.
If not, $[\mathcal J^a(\ell),\chi(x')]=0$.
If $\ell(x,e_i)$ is on $\mathcal P(x')$, the path factorizes as $W(x')=W(x)\,U(x,e_i)\,W(x+e_i\to x')$, and
\begin{align}
[E^a_L(\ell),\chi(x')]
=W(x)\,T^a\,U(x,e_i)\,W(x+e_i\to x')\,\psi(x')
=\left[W(x)\,T^a\,W(x)^\dagger\right]\chi(x')\,.
\end{align}
Contracting with $W^1_{ab}(x)$ removes the transporter, since $W^1_{ab}\,W\,T^b\,W^\dagger=T^a$, and
\begin{align}
\label{sm:Jchi}
[\mathcal J^a(\ell),\chi(x')]=T^a\,\chi(x')\,,\qquad
[\mathcal J^a(\ell),\chi^\dagger(x')]=-\chi^\dagger(x')\,T^a\,.
\end{align}
Let $S(\ell)$ be the piece of the tree beyond $\ell$, the sites whose path $\mathcal P$ passes through $\ell$.
Comparing Eq.~\eqref{sm:Jchi} with Eq.~\eqref{sm:Qpsi}, $\mathcal J(\ell)$ acts on the fermions as minus the total charge beyond $\ell$,
\begin{align}
\label{sm:QS}
Q^a_{S(\ell)}=\sum_{x'\in S(\ell)}\chi^\dagger(x')\,T^a\,\chi(x')\,,
\end{align}
with no rod multiplying it, for any maximal tree in any dimension.

\begin{figure}[t]
\centering
\includegraphics[width=0.9\columnwidth]{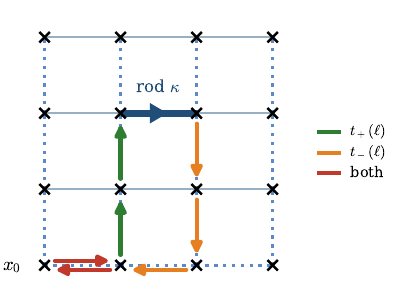}
\caption{The comb maximal tree on a $4\times4$ patch of the boundary, with the origin $x_0$ at the lower left, tree links dotted and rods solid.
The loop $P(\kappa)$ of one rod $\kappa$ runs from the origin along the tree, across $\kappa$, and back to the origin along the tree.
Each tree link on the loop is colored by the set of Eq.~\eqref{eq:J} that $\kappa$ belongs to for that link. The loop traverses it outward only ($t_+$), on the return only ($t_-$), or in both directions, which happens on the stretch shared by $\mathcal P(x)$ and $\mathcal P(x+e_i)$.}
\label{fig:tree}
\end{figure}

Let $t_+(\ell)$ and $t_-(\ell)$ be the sets of physical links $\kappa$ whose loop $P(\kappa)$ traverses $\ell$ in the positive and in the negative direction (Fig.~\ref{fig:tree}).
A loop traverses $\ell$ in both directions when $\ell$ lies on both tree paths, and such a $\kappa$ belongs to both sets.
A positive traversal contributes the left generator of the rod and a negative one minus its right generator~\cite{DAndrea:2023qnr}, so with Eq.~\eqref{sm:Jchi}
\begin{align}
\label{sm:J}
\mathcal J^a(\ell)=\sum_{\kappa\in t_+(\ell)}\mathcal E^a_L(\kappa)-\sum_{\kappa\in t_-(\ell)}\mathcal E^a_R(\kappa)-Q^a_{S(\ell)}\,,
\end{align}
which is Eq.~\eqref{eq:J}. For a physical link the charge term is absent, since the paths that define the transported fermions contain tree links only.
Eq.~\eqref{sm:J} is an identity on the states that satisfy Gauss's law at every site other than the origin, whatever their total color. Those states are generated from one reference state, with no electric flux and no fermions, by the rods and the transported fermions~\cite{Kogut:1974ag}. Both sides of Eq.~\eqref{sm:J} annihilate that state and have the same commutators with every rod and every transported fermion, so they agree on every state generated from it.

Since $W^1$ is orthogonal, $E^2_\ell=\mathcal J^2_\ell$, and the electric Hamiltonian is the first line of Eq.~\eqref{eq:H} with $\Phi=0$.
The fermions enter through the total color charge beyond each tree link, a long-range interaction along the tree of the same kind as the one among the rods. The magnetic Hamiltonian is unchanged.

\subsection{The global constraint}
\label{app:global}
Gauge fixing solves Gauss's law identically at every site other than the origin.
What remains is the transformation at the origin, which acts on the rods and on all the fermions at once,
\begin{align}
X_\kappa\to\Omega(x_0)\,X_\kappa\,\Omega(x_0)^\dagger\,,\qquad \chi(x)\to\Omega(x_0)\,\chi(x)\,,
\end{align}
generated on the rods by $\sum_\kappa[\mathcal E^a_L(\kappa)-\mathcal E^a_R(\kappa)]$ and on the fermions by $-\sum_xQ^a(x)$.
With
\begin{align}
L^a_\text{tot}=\sum_\kappa\left[\mathcal E^a_R(\kappa)-\mathcal E^a_L(\kappa)\right],\qquad
Q^a_\text{tot}=\sum_xQ^a(x)\,,
\end{align}
the one constraint the gauge fixing leaves is Eq.~\eqref{eq:global}. The rods may carry color provided the fermions carry the opposite color.
$L_{\rm tot}$ acts on the rods by conjugation and is blind to the center of $G$, which is why it is integer for \sutwo{}.

\subsection{Fermions in the bulk}
\label{sm:bulk}
Write $\Lambda$ for the slab, with boundary sites $B$ and bulk sites $I$ and a fermion $\psi(y)$ on every site.
Only the links with both ends on the boundary are dynamical.
The links with an end in the bulk have no conjugate momenta. Their values are the functions $F_\ell[U_B]$ of the boundary links, Eq.~\eqref{eq:flow}, where $U_B$ is the set of boundary links before gauge fixing and $F_\ell[U_B]$ is the bulk link $\ell$ of the configuration $F^{(n)}$ of the main text.
The fermion Hamiltonian is the Wilson Hamiltonian of the domain-wall fermion on all of $\Lambda$, with the color matrix on each link equal to $U_\ell$ for a boundary link and to $F_\ell[U_B]$ otherwise.
The maximal tree of the main text lies on the boundary, and gauge transformations act on the boundary sites only. Before that gauge fixing can be carried out on the slab, two things have to be fixed, how the bulk links and the bulk fermions transform.

Under a boundary gauge transformation $\{\Omega(x)\}_{x\in B}$ the bulk links change to $F_{yy'}[\Omega\cdot U_B]$.
The bulk hopping terms are gauge invariant if and only if this change is a gauge transformation of the bulk link at its two ends,
\begin{align}
\label{sm:Fcov}
F_{yy'}\big[\Omega\cdot U_B\big]=\tilde\Omega(y)\,F_{yy'}\big[U_B\big]\,\tilde\Omega(y')^\dagger\,,
\end{align}
for some $\tilde\Omega(y)\in G$ at every bulk site. The same $\tilde\Omega(y)$ acts on every link ending at $y$, and $\tilde\Omega(x)=\Omega(x)$ on the boundary. The bulk fermion then transforms as $\psi(y)\to\tilde\Omega(y)\psi(y)$.
Eq.~\eqref{sm:Fcov} holds for the bulk of Eq.~\eqref{eq:flow}, since the action is gauge invariant and the far-wall prescription is covariant, so the stationary point over the transformed boundary is a bulk gauge transform of the stationary point over the original one.
A gauge condition in the bulk selects one representative, and $\tilde\Omega$ is the bulk transformation that restores that condition after $\Omega$ has acted.
For a general bulk gauge condition $\tilde\Omega$ depends on the boundary field as well as on $\Omega$. The condition of the main text removes that dependence.

The bulk gauge condition is that every link along $s$ is the identity.
It is complete, since the links along $s$ from a boundary site form a path to each bulk site, and the transformation that brings a configuration to this form is unique and trivial on the boundary.
Restoring the condition after a boundary transformation carries $\Omega(x)$ along $s$,
\begin{align}
\label{sm:attach}
\tilde\Omega(x,s)=\Omega(x)\,,
\end{align}
independent of the boundary field.
Gauss's law at a boundary site then counts the charge of every fermion along $s$ from it,
\begin{align}
\label{sm:gauss_bulk}
\Big[G^a(x)+\sum_{s\ge0}Q^a(x,s)\Big]\ket{\Psi_\text{phys}}=0\qquad\forall\,x\in B\,,
\end{align}
and a bulk link from $(x,s)$ to $(x',s)$ transforms like a boundary link from $x$ to $x'$.
Any other bulk gauge condition is reached from this one by a bulk transformation $h(y)[U_B]$, trivial on the boundary, that depends on the boundary field.
It is a change of fermion variables, $\psi(y)\to h(y)\psi(y)$, and gives the same Hamiltonian conjugated by the unitary $\exp\big(i\sum_y\psi^\dagger(y)\log h(y)\,\psi(y)\big)$, provided the electric operators, which differentiate $h[U_B]$, are shifted by the bulk charges dressed with that derivative.
The condition along $s$ used here is the one in which no such shift appears, so the electric operators keep the form of Eq.~\eqref{sm:J} and Gauss's law at a boundary site counts the charges along $s$ from it, Eq.~\eqref{sm:gauss_bulk}. Every other bulk gauge is unitarily equivalent to it, at the price of electric operators dressed with derivatives of the bulk transformation.

The gauge fixing on the boundary is now carried out as before.
A bulk fermion at $(x,s)$ is transported to the origin along the tree from the boundary site $x$, $\chi(x,s)=W(x)\,\psi(x,s)$, and the bulk hopping terms become
\begin{align}
\label{sm:hop_bulk}
\mathsf h(y,y')=\chi^\dagger(y)\,\Gamma\,F_{yy'}(X)\,\chi(y')\,,\qquad
F_{yy'}(X)=W(x)\,F_{yy'}\big[U_B\big]\,W(x')^\dagger\,,
\end{align}
for $y=(x,s)$ and $y'=(x',s')$.
By Eq.~\eqref{sm:Fcov}, $F_{yy'}(X)$ is invariant under every gauge transformation that is trivial at the origin, so it is a function of the rods alone, equal to $F_{yy'}$ on the gauge-fixed boundary, and under the transformation at the origin it is conjugated, $F_{yy'}(hXh^{-1})=h\,F_{yy'}(X)\,h^{-1}$, which is Eq.~\eqref{eq:equivariant}.
The commutators of $\mathcal J(\ell)$ with $\chi(x,s)$ are those with $\chi(x)$, so Eq.~\eqref{sm:J} holds with the charge beyond a tree link including the fermions along $s$,
\begin{align}
\label{sm:QS_bulk}
Q^a_{S(\ell)}=\sum_{x\in S(\ell)}\sum_{s\ge0}\chi^\dagger(x,s)\,T^a\,\chi(x,s)\,,
\end{align}
and $Q_{\rm tot}$ in Eq.~\eqref{eq:global} counts every fermion, boundary and bulk.
The hopping terms do not know at which boundary site a bulk fermion is counted. The electric Hamiltonian does, through Eq.~\eqref{sm:QS_bulk}, since the charge of a bulk fermion at $(x,s)$ sources the boundary electric field as a charge at $x$ would.
Counting it at a different boundary site $x'$ would add a string of boundary electric flux from $x$ to $x'$, a different Hamiltonian rather than a different gauge condition, and on the slab the choice along $s$ is the only translation-invariant one.

\subsection{The orientation of the bulk instanton}
\label{sm:orientation}
A boundary configuration is reducible if some $h\in G$ outside the center commutes with every rod.
The trivial boundary $X=1$ is reducible under all of $G$, and a boundary whose rods lie in one $U(1)$ subgroup is reducible under that $U(1)$.
At a reducible boundary Eq.~\eqref{eq:equivariant} constrains the bulk. At $X=1$ every bulk link would have to lie in the center, a flat bulk with no charge, and at a boundary whose rods lie in one $U(1)$ subgroup the orientation of the bulk instanton in color space is fixed only up to rotations about that axis.
Over $X=1$ the orientation is a zero mode. A global rotation of the bulk leaves the identity boundary and the flat wall unchanged, so the stationary points of sector $n$ form the orbit of one lump under $G$, and over a boundary within $\delta$ of the trivial one every stationary point lies within $O(\delta)$ of that orbit.
Over $X=1$ the lump's transverse position is a translation orbit as well, and the seed's center selects one. Both are sets of boundaries of measure zero, and what a rod trajectory does across them belongs to the dynamics not computed here.
A single-valued $F^{(n)}$ satisfying Eq.~\eqref{eq:equivariant} on a punctured neighborhood of $X=1$ would assign the lump an orientation continuously, rotating with the rods, and no such assignment exists.
Near $X=1$ the irreducible configurations form a simply connected set. In it take the loop of rods $h(t)Xh(t)^{-1}$, with $h(0)=1$ and $h(1)$ a nontrivial element of the center. The loop is closed, since $h(1)$ commutes with $X$, and contractible, since the set is simply connected. By Eq.~\eqref{eq:equivariant} the lump's orientation would follow $h(t)$, a loop in $G/Z(G)$ that winds once around a generator of $\pi_1(G/Z(G))=Z(G)$ and is not contractible. A continuous map cannot send a contractible loop to a non-contractible one, so no such assignment exists.
The flow is covariant, so $F^{(n)}$ continued along any path of rods satisfies Eq.~\eqref{eq:equivariant} along that path. What fails is single-valuedness.

The definition of the main text takes the stationary point of lowest action.
That choice is unique except on the sets where two orientations have equal action, which we call tie sets. They have codimension one and emanate from the reducible set.
Across a tie set the lump rotates in color space, the action is continuous and the state of the fermion sea is not.
Near the reducible set the action depends only weakly on the orientation and the tie sets are dense there. At a generic boundary the dependence is of order one and the minimum is unique.
The Hamiltonian is self-adjoint regardless, since the bulk enters it only through bounded fermion bilinears and the tie sets have measure zero.
A rotation of the lump is a reorganization of the bulk within a sector, of the same kind as the change of label at the sphaleron. The boundary sees it as it sees the lump, through the zero modes of the covariant Laplacian and through tails of order $e^{-ks}$, so it moves the boundary by $O(\delta)$ in the zero modes and by exponentially little otherwise. The excitation it leaves in the bulk fermions is gapped and localized.
At the reducible set itself the orientations are gauge copies.
For \sutwo{} that set is where every rod's rotation axis is the same, of codimension two for every rod beyond the first, and there the Haar measure of the relative orientations vanishes and the kinetic term of the mixed basis carries a centrifugal barrier in the relative axis angular momentum, the many-rod form of the single-plaquette barrier of Ref.~\cite{DAndrea:2023qnr}.

\subsection{The screened electric coupling}
\label{sm:screened}
The shift $D_X\Phi$ in Eq.~\eqref{eq:H} is built from the covariant difference on adjoint fields of the boundary,
\begin{align}
\label{sm:DX}
(D_X\Phi)^a_\ell={\rm Ad}(U_\ell)^{ab}\,\Phi^b_y-\Phi^a_x\,,\qquad {\rm Ad}(U)^{ab}=\tfrac12\Tr\big(\sigma^aU\sigma^bU^\dagger\big)\,,
\end{align}
for $\ell$ from $x$ to $y$, with $U_\ell$ the rod or the identity, and $\Delta_X=D_X^{T}D_X$.
Under the residual global $G$, and under a local transformation in the unfixed form, $D_X\Phi$ transforms as $\Phi$ does, so $\mathcal J-D_X\Phi$ transforms as $\mathcal J$ and Eq.~\eqref{eq:global} is unchanged.
The shifted field obeys
\begin{align}
\label{sm:screenedgauss}
D_X^{T}\big(\mathcal J-D_X\Phi\big)=Q_{\rm bdy}+\sum_{s\ge1}f_s(\Delta_X)\,Q_s\,.
\end{align}
This is Gauss's law for the shifted flux. The boundary charge enters in full and the charge on layer $s$ enters through $f_s(\Delta_X)$, in full at zero momentum and reduced otherwise, the form factor of \EM{em:coupling}{A}. Its square is the electric term.

\paragraph{The choice of kernel.}
The profile $f_s$ of Eq.~\eqref{eq:Phi} is the extension of the Gauss-law multiplier into the bulk, and any extension with $f_0=1$ is allowed.
Solving the Laplace equation in the bulk, as in Ref.~\cite{Sen:2026hfw}, spreads a far charge over a transverse range of order $L_s$, while the heat kernel of gradient flow, $f_s(\mu)=e^{-s\mu}$, has range $\sqrt{2s}$.
The range sets how many rods one term acts on at once, so the shorter range is the practical choice.
The coupling is local relative to the box when $L_b^2\gg L_s$, and a flow time capped at $\tau^*$, $f_s(\mu)=e^{-\min(s,\tau^*)\mu}$, makes the range independent of $L_s$.
Decoupling needs $aL_s\to\infty$ while the range $a\sqrt{L_s}$ must vanish, so $L_s$ is taken between $1/a$ and $1/a^2$. The elliptic kernel's range is the depth $aL_s$ itself, so for it no such window exists.
A change of kernel changes how the bulk charges enter the electric term and nothing else, so different kernels are different regulated theories and the difference lives in the bulk.
The boundary is insensitive to the choice because the sheet difference is the response of gapped charges at depth $s$, which a damped kernel weights by $f_s(k^2)$ at transverse momentum $k$, so two damped kernels differ by terms that fall with the depth of the charge.
The point coupling $f_s=1$ lets the mirror see the boundary field undamped, which is why it fails.
The subtracted electric sheet difference is below $10^{-4}$ for the heat-kernel, Gaussian and elliptic profiles (Appendix~\ref{sm:pvchecks}).

\subsection{The register}
\label{sm:register}
The configuration space of the rods and the register is the covering space of rod space with the exceptional set of Eq.~\eqref{eq:NCS} removed, on which a winding is not a closed path (\EM{em:register}{B}).
A point of it is a rod configuration together with its Chern-Simons number $\NCS[X]$ of Eq.~\eqref{eq:NCS} followed continuously from the trivial boundary, and the sheets of the main text are the sheets of this cover.
The Hilbert space is the space of wavefunctions on the cover. On sheet $n$ the wavefunction is a function $\psi_n(X)$ of the rods, with the Haar measure, and its values lie in the one Fock space of the slab fermions. The Hamiltonian acts on sheet $n$ with the bulk $F^{(n)}$.
No cut is made in the space. A representation stores $\psi_n(X)$ with $X$ in a fundamental domain and $n$ in a register, the domain's edge is the sphaleron surface where $n$ changes by one, and two such choices differ by a relabeling.
A plaquette dislocation, a step on which L\"uscher's interpolation of a plaquette crosses its cut, jumps the integral of Eq.~\eqref{eq:NCS} by an integer without changing $\NCS$ on its branch or the register.
Every loop that changes the register passes around the set on which $\NCS$ mod 1 is undefined, and such a loop can be short, a path through fields with a plaquette far from the identity. Those fields cost a magnetic energy of order $1/g^2$ per such plaquette, so at weak coupling they are suppressed as dislocations are, and an admissibility condition on the boundary would exclude them. Their rate is not computed here.
A phase $e^{i\theta}$ in the matching across the sphaleron surface is removed by the unitary $e^{i\theta\hat n}$, with $\hat n$ the register, which acts on the boundary as the rotation by its fermion number, so there is no vacuum angle to choose, for any content with a Weyl fermion. Without fermions the sheets are exactly degenerate and $\theta$ returns as the Bloch label of the shift $n\to n+1$. The matching phase drops out of the spectrum and of the neutral boundary correlators, and the sectors of net winding enter boundary correlators only through insertions that carry net chirality, whose phase the relabeling of the sheets absorbs.
The Pauli-Villars connection of Appendix~\ref{app:pv} lives on the same cover.
The register is bounded only by the bulk's capacity, of order the number of bulk plaquettes times the $\epsilon$ of Appendix~\ref{sm:bulkcomp}.
The inverse winding is the charge conjugate of a winding, a level rising from the filled band onto the level the anti-instanton binds at $+|\epsilon_0|$.

\section{Pauli-Villars partners in the Hamiltonian}
\label{app:pv}
The bulk fermions have a vacuum energy, a potential for the Polyakov loops and a charge density that responds to the boundary field, all of which grow with $L_s$.
In the Euclidean domain-wall theory these are removed by Pauli-Villars fields, bosonic spinors with the same operator and mass as the fermions, whose determinant divides that of the fermions~\cite{Furman:1994ky,Vranas:1997da}.
Without them the energy difference between two sheets would contain the bulk's own response to the bulk instanton, of order the bulk mass, and would grow with $L_s$.
Fields that enter the Euclidean theory only through an inverse determinant cannot enter a Hamiltonian as fields. A Grassmann-even field with an inverse determinant requires a Fock space of indefinite norm, which light-front treatments accept as a regulator artefact and project out at the end of a calculation~\cite{Brodsky:1998hs,Paston:1999fp}. That is possible for amplitudes and not for a state evolved unitarily, which is what a quantum simulation does.
Instead, here the partners enter as functions of the rods.

\subsection{The Pauli-Villars operator}
\label{sm:pvop}
\begin{figure}[t]
\centering
\includegraphics[width=0.95\columnwidth]{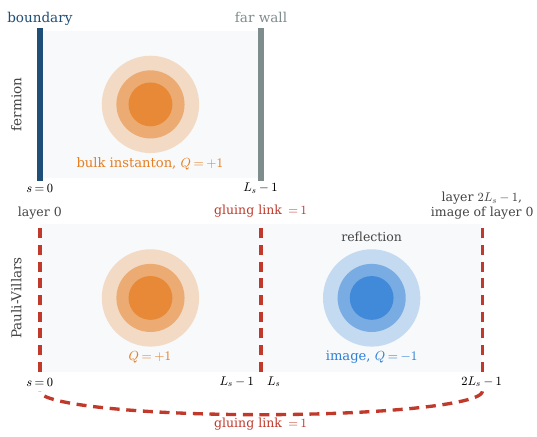}
\caption{The Pauli-Villars closure. Top, the fermion slab of Fig.~1(a), layers $s=0$ to $L_s-1$, with the bulk instanton. Bottom, the closed lattice of the partners, the slab followed by its reflection, layers $s=0$ to $2L_s-1$, with the instanton's image of opposite charge. The gluing links carry the identity, and the closed lattice has no open end and so no chiral modes.}
\label{fig:closure}
\end{figure}

Let $h_f[X]$ be the single-particle Wilson operator of the bulk fermions on the slab, in the background set by the rods $X$.
The Pauli-Villars operator $h_{\rm PV}[X]$ is the same Wilson operator, with the same $M$ and the same color matrices on the links parallel to the boundary, on a closed lattice, the slab followed by its reflection, $T^3\times S^1$ of length $2L_s$ in $s$, with layer $L_s+j$ a copy of layer $L_s-1-j$ (Fig.~\ref{fig:closure}).
The links along $s$ are the identity, the field is antiperiodic around the closed direction, and its functionals enter with weight one half.
This is the closure of Ref.~\cite{Kikukawa:2001mw}.
The reflection reverses the orientation of $s$, so the closed lattice carries the bulk instanton and an image of opposite charge, with total charge zero, and the field is smooth across both gluing links. We call this the link closure, since the slab and its reflection are joined across links. The alternative, in which the reflection is taken about the first and last layers so that each appears once on a circle of $2L_s-2$ layers, we call the layer closure (Appendix~\ref{sm:pvdyn}).
The closed lattice has a particle-hole symmetry that the slab's chiral modes break, so its spectrum is exactly symmetric about zero while the slab's spectrum is not. The bound level at $\epsilon_0$ has no partner on the slab and has one on the closed lattice.
The Pauli-Villars sea therefore has exactly half of its levels filled on every sheet and no net spectral flow along a winding. Its functionals are continuous in the rods, with at most a kink where a pair of levels at $\pm\varepsilon$ reaches zero.
The two levels the closed lattice binds, one to the instanton and one to its image, hybridize into such a pair, the $\pm$ pair below.
The subtraction removes the bulk's local response to the bulk instanton, which the closed lattice sees twice over, and not the instanton's topological charge, which it does not see at all.
The other closure, the slab closed on itself with one antiperiodic seam from the far wall back to the boundary, is the one of the transfer-matrix treatments of domain-wall fermions~\cite{Neuberger:1997bg,Vranas:1997da}.
It would join a flat wall to a boundary that is not flat, and the partners would couple the boundary to the mirror through the seam~\cite{Ago:2019yjt}, with a contribution to the subtracted energies that does not vanish as $L_s$ grows. The link closure, and the layer closure with it, joins each layer to its own image instead.

\subsection{Static backgrounds}
\label{sm:pvstatic}
Fix the rods, so that every color matrix $F_\ell(X)$ is a number, and consider the bulk fermion part of the Hamiltonian,
\begin{align}
H_f[X]=\psi^\dagger\,h_f[X]\,\psi\,.
\end{align}
Its Euclidean partition function at inverse temperature $\beta$ is the determinant of the time-discretized Wilson operator in the static background, and the transfer-matrix construction~\cite{Creutz:1976ch,Luscher:1976ms} gives
\begin{align}
\det D_f[X]={\rm Tr}\,e^{-\beta H_f[X]}
\ \xrightarrow{\ \beta\to\infty\ }\
e^{-\beta E_f[X]}\,,\qquad
E_f[X]=\sum_{w^f_k[X]<0} w^f_k[X]\,,
\label{eq:pv_tm_f}
\end{align}
with $w^f_k[X]$ the eigenvalues of $h_f[X]$ and $E_f$ the energy of the filled Dirac sea.
The Pauli-Villars field contributes the inverse determinant of the same kind of operator, and in a static background the same projection applies to it with the opposite sign in the exponent,
\begin{align}
(\det D_{\rm PV}[X])^{-1}\ \xrightarrow{\ \beta\to\infty\ }\
e^{+\beta E_{\rm PV}[X]}\,,\qquad
E_{\rm PV}[X]=\tfrac12\sum_{w^{\rm PV}_k[X]<0} w^{\rm PV}_k[X]\,,
\label{eq:pv_tm_pv}
\end{align}
with $w^{\rm PV}_k[X]$ the eigenvalues of $h_{\rm PV}[X]$ and the factor one half from the closure.
The ground-state energy of the subtracted theory in a static background is therefore
\begin{align}
E[X]=E_f[X]-E_{\rm PV}[X]=E_f[X]+V^{\rm PV}[X]\,,\qquad V^{\rm PV}[X]\equiv-E_{\rm PV}[X]\,,
\label{eq:pv_static}
\end{align}
which is the same quantity as the energy of the Euclidean domain-wall theory with its Pauli-Villars fields in a static background, since for a static field that path integral is $e^{-TE}$ with $E$ this difference.
In the static sector the partners contribute only a potential on rod space. It is a function of the rods alone, because the bulk links are functions of the rods too, and it depends on the sheet through $F^{(n)}$. The connection and the inertia correction below appear only when the rods move.
Every number in this Letter is a static-sector quantity, so the subtraction behind every number is Eq.~\eqref{eq:pv_static}.

\subsection{Beyond the static sector}
\label{sm:pvdyn}
When the rods are dynamical, the partners' determinant depends on the path $X(t)$ and not only on its endpoints. Whether the partners can still be written as functions of the rods depends on how fast the rods move.
A rod mode is a normal mode of the rods about a background, with frequency $\nu^2\sim V''/\mathcal G$ set by the electric and magnetic terms, $\mathcal G$ being the rods' inertia of Eq.~\eqref{eq:Ginertia} below, and $\nu\approx|k|$ in the weak-field limit for transverse momentum $k$.
A mode is slow if $\nu$ is small compared with the gap of $h_{\rm PV}$, which is of order $M$, and fast if $\nu$ is comparable to or above it. The modes of long wavelength and the Polyakov loops are slow. The ultraviolet modes are fast.

For the slow modes, integrating the Pauli-Villars sea out adiabatically gives its effective action for the rods as an expansion in $\nu/M$. The first three orders are
\begin{align}
S^{\rm PV}_{\rm eff}[X(t)]&=\int dt\,\Big[-V^{\rm PV}[X]+\sum_\kappa A^{\rm PV}_\kappa[X]\,\dot X_\kappa
+\tfrac12\sum_{\kappa\kappa'}\dot X_\kappa\,\mathcal I^{\rm PV}_{\kappa\kappa'}[X]\,\dot X_{\kappa'}+O\big((\nu/M)^3\big)\Big],
\label{eq:pv_bo}\\
A^{\rm PV}_\kappa[X]&=-\,i\sum_{w_k<0}\langle k[X]|\partial_\kappa\,k[X]\rangle,
\qquad
\mathcal I^{\rm PV}_{\kappa\kappa'}[X]=-\,2\,{\rm Re}\sum_{w_k<0<w_m}
\frac{\langle k|\partial_\kappa h_{\rm PV}|m\rangle\langle m|\partial_{\kappa'}h_{\rm PV}|k\rangle}{(w_m-w_k)^3},
\label{eq:pv_AM}
\end{align}
with $|k[X]\rangle$ the eigenvectors of $h_{\rm PV}[X]$, $\partial_\kappa$ the left derivative with respect to the rod $X_\kappa$, the direction generated by $\mathcal E_L(\kappa)$, $\dot X_\kappa$ the corresponding angular velocity, and the overall signs opposite to those of the fermion sea.
Order zero is the potential of Eq.~\eqref{eq:pv_static}.
Order one is a Berry connection on rod space, which shifts the rod generators, $\mathcal E_L(\kappa)\to\mathcal E_L(\kappa)-A^{\rm PV}_\kappa[X]$, and of which only the curvature is physical.
A connection on a space with a non-contractible loop can carry a holonomy around that loop, which no redefinition of the connection removes. In a pure gauge theory a winding is such a loop and the holonomy is the vacuum angle. Here the configuration space is the cover of Appendix~\ref{sm:register}, on which a winding is not a closed path, so the winding is not a loop for the connection to carry a holonomy around.
Order two is a polarizability, a negative correction to the inertia of the rods. It corrects the electric coefficient alone, and since on a Hamiltonian lattice the electric and magnetic coefficients are independent and their ratio sets the speed of light, it is a renormalization of that ratio.

Eq.~\eqref{eq:pv_bo} is a Lagrangian for the rods, local in time, and its Legendre transform is a Hamiltonian, with operator-ordering terms of the order the expansion drops.
For the fermion sea the same three terms are the Born-Oppenheimer expansion, the sea's ground-state energy, Berry connection and quantum geometric tensor, which agree with the derivative expansion of the fermion determinant order by order. The partners' functionals are the same spectral formulas with the regulator's sign.
The Hamiltonian of the rods is then
\begin{align}
H_{\rm reg}=H\Big|_{\mathcal E_L(\kappa)\to\mathcal E_L(\kappa)-A^{\rm PV}_\kappa,\ \ \mathcal G\to\mathcal G+\mathcal I^{\rm PV}}+V^{\rm PV},
\label{eq:Hreg_sm}
\end{align}
Eq.~\eqref{eq:Hreg} of \EM{em:pv}{C}, with the rod generators shifted in the electric term of Eq.~\eqref{eq:H}, the charge terms and $D_X\Phi$ unchanged, and $\mathcal G$ the inertia of the rods, the quadratic form in the rod generators of that term,
\begin{align}
\mathcal G^{-1}=g^2\big(1+T^{\rm T}T\big)\,,
\label{eq:Ginertia}
\end{align}
where $T$ writes the electric field of each tree link as the sum of the rod fields whose loops traverse it, Eq.~\eqref{eq:J}. Eq.~\eqref{eq:H} contains $\mathcal G$ already, since its electric term is $\tfrac12 E^{\rm T}\mathcal G^{-1}E$ in the rod generators $E$. The partners shift that mass matrix.
$H_{\rm reg}$ is bounded from below only where $K=\mathcal G+\mathcal I^{\rm PV}>0$.

A change of the rods changes the bulk links, through $F^{(n)}(X)$, and the electric fields on the tree links, through $T$. The inertia correction $\mathcal I^{\rm PV}$, the partner sea's quantum geometric tensor of \EM{em:pv}{C}, is the sum of a transverse part from the first and a Coulomb part from the second.
The transverse part is the partner sea's response to the bulk field. For the Polyakov loops, whose change reaches every layer, it adds up over the slab and grows linearly with $L_s$, while for a mode of transverse momentum $k$ the bulk field follows the rod only to depth $1/k$ and the response is bounded.
The Coulomb part is the response to the electric field along the tree, which extends through the slab. It is also extensive in $L_s$, and it acts on the rod combinations whose Coulomb energy is highest and whose inertia $\mathcal G$ is therefore smallest. Without gauge fixing it is the longitudinal polarizability of the partner sea, and the fermion sea has the same polarizability with the opposite sign, the dielectric of the physical picture. The condition $K>0$ quantifies whether the dielectric constant of the bulk, fermions and partners together, stays positive.
On the volumes computed here, the transverse part alone would violate $K>0$ only at $g^2L_s$ of order $10^2$, and the Coulomb part violates it at $g^2L_s$ of order a few (Appendix~\ref{sm:results}).
The modes where it fails are stiff, with $\nu^2\sim V''/\mathcal G$ several times the bulk gap, outside the expansion in $\nu/M$ that defines $\mathcal I^{\rm PV}$.
Above that coupling Eq.~\eqref{eq:Hreg} is unbounded from below, a negative inertia that no fermion term offsets, so the three-functional form is a Hamiltonian for the slow rod modes only below that coupling.
For the fast ones the fermion sea responds to the same fields with the opposite sign, and the two responses are to be integrated out together. The functional we have in mind is built from the deep bulk of both seas, the modes above a threshold inside the gap, and the light modes kept as fields. Its inertia is $\mathcal I^f+\mathcal I^{\rm PV}$ summed over the deep modes only. Computed for the two full seas, with no threshold, $\mathcal I^f+\mathcal I^{\rm PV}$ is bounded and small in every direction, and the response of the two seas together is a few percent of either one and independent of $L_s$ (Appendix~\ref{sm:results}). On $3^3$ and $4^3$, where the fermion inertia was computed, the slab's lightest level lies above $1.2$, so no mode sits inside the gap and the two are the same functional. On larger volumes the boundary bands enter the gap and stay fields, and their removal from the functional is not computed.
In the gauge-fixed metric $\mathcal G+\mathcal I^f+\mathcal I^{\rm PV}>0$ then holds for $g^2$ below $33$, $32$ and $31$ at $L_s=4$, $8$ and $12$ on $3^3$ and below $12$ and $11$ at $L_s=4$ and $6$ on $4^3$, where $\mathcal G+\mathcal I^{\rm PV}>0$ alone holds only below $1.3$, $0.7$ and $0.5$ on $3^3$ (Appendix~\ref{sm:results}).
The residue that sets the bound is a surface term. The link closure of Appendix~\ref{sm:pvop} reflects the slab across a link, so the first and last layers each appear twice, and the residue is negative. The layer closure reflects about those layers, each once on a circle of length $2L_s-2$ at the same weight, so the partners cover $L_s-1$ layers and half a layer of the fermion sea is left unsubtracted at each end, a positive surface term. The inertia of both seas is then positive definite, with lowest eigenvalue $+0.0008$ at $L_s=4$ and $8$ on $3^3$ and $+0.0012$ on $4^3$, and $K>0$ for every $g^2$.
The bulk cancellation is exact for either closure, and the choice between them is a regulator choice at the ends of the slab.
The light modes, the boundary and mirror bands and the level bound to the instanton, stay fields, so the crossing stays explicit.
The functionals of the integrated-out sector are traces of rational functions of the single-particle operators, as in the overlap construction, and need no diagonalization. That scheme, with the test that the result does not depend on the threshold, is left for future work.

A Hamiltonian with a positive-norm Hilbert space always has a reflection-positive Euclidean transcription. The Euclidean domain-wall action with its Pauli-Villars fields is not reflection positive, since a bosonic spinor field with a local action has no positive transfer matrix. In the transfer-matrix derivations of the overlap~\cite{Neuberger:1997bg,Vranas:1997da} the inverse determinant is the norm of a boundary state, not the propagation of a field, which is the role it plays here as well.
A Hamiltonian with a positive-norm Hilbert space therefore cannot reproduce that action exactly.
The Hamiltonian with the functionals is a different regulator of the same slab theory, with a positive-norm Hilbert space. It agrees with the Pauli-Villars-subtracted Euclidean slab theory exactly in its static sector and, for the slow rod modes, up to irrelevant operators. The statement is about the slab as a whole, a vector-like theory, which is what the Euclidean domain-wall theory with Pauli-Villars fields regulates~\cite{Furman:1994ky,Vranas:1997da}. What the boundary alone is, in either regulator, is the question of Sec.~\ref{sec:anomalies}.

\section{The far wall}
\label{app:wall}
The far wall is a flat configuration, every plaquette on it equal to one.
A flat configuration on a three-torus is fixed by its three commuting holonomies, which the construction sets from the boundary's three Polyakov loops.
The numerical checks are in Appendix~\ref{sm:wallchecks}.

\subsection{Prescription}
\label{sm:wallrule}
The boundary's Polyakov loops need not commute. They fail to commute whenever the boundary carries net flux through a plane, and then no flat wall has the same holonomies, and no continuous prescription assigns a flat wall to every boundary. A wall with 't Hooft twisted boundary conditions would carry non-commuting holonomies, at the price of a twist in the bulk, and is not used here.
Ours reads the Polyakov loops at every boundary site, averages them, and projects the average onto their principal axis, as in \EM{em:wall}{D}.
For \sutwo{}, let
\begin{align}
W_i(x)=\exp\big(i\,\phi_i(x)\,\hat\eta_i(x)\cdot\sigma\big)
\end{align}
be the product of the boundary links around the cycle in direction $i$ from site $x$, transported to the origin along the tree. On the gauge-fixed boundary the transport is trivial.
Average over the sites,
\begin{align}
\bar b_i=\big\langle\sin\phi_i\,\hat\eta_i\big\rangle_x\,,\qquad \bar c_i=\big\langle\cos\phi_i\big\rangle_x\,,
\end{align}
let $\hat\eta$ be the principal axis of the three vectors $\bar b_i$, the unit vector that maximizes $\sum_i(\bar b_i\cdot\hat\eta)^2$, and set
\begin{align}
W^{\rm wall}_i=\exp\big(i\,\varphi_i\,\hat\eta\cdot\sigma\big),\qquad \varphi_i=\mathrm{atan2}\big(\bar b_i\cdot\hat\eta,\,\bar c_i\big)\,.
\label{eq:wallks_sm}
\end{align}
This is the projection of the averaged loop onto the $U(1)$ generated by $\hat\eta\cdot\sigma$.
For a flat boundary every $W_i(x)$ is the same element, the loops commute, and the wall carries them, including $W_i=-1$.
The prescription needs only the Polyakov loops up to a common conjugation, so it can be stated for a boundary in Kogut-Susskind variables as well, with the links along $s$ set to the identity as in any domain-wall construction.

\subsection{Properties}
\label{sm:wallprops}
Under a gauge transformation $h$ at the origin every loop is conjugated by $h$, the averaged vectors and the axis rotate with them, and the wall holonomies are conjugated by the same $h$ carried along $s$, so the prescription is covariant.
It depends on no single base point. A prescription that reads the loops at one site does, since loops based at different sites of a boundary with flux are not conjugate. The loops are compared after transport along the tree. For a flat boundary, where all $W_i(x)$ coincide, the transport drops out.
It is continuous except where the axis $\hat\eta$ is undefined. That happens at boundaries whose three vectors $\bar b_i$ have no single principal direction, a set of codimension two in the space of Polyakov loops, which we call the singular set of the prescription. The points where $\bar c_i=0$ and $\bar b_i\perp\hat\eta$, so that $\varphi_i$ is undefined, belong to the same set.
A discontinuity of the wall could leave a mark on the fermion sea as a Berry phase around a loop in rod space that encircles the singular set. Appendix~\ref{sm:wallchecks} finds no such phase on the loops computed.

\subsection{Tension with the bulk}
\label{sm:wallmismatch}
The wall is set by the boundary alone, and the wound bulk's action is not stationary in the wall's holonomy. Rotated about the instanton's axis at fixed boundary, the action falls, and if the wall's holonomies are allowed to move under the flow they settle away from the prescription's value, with the action lower by a few percent and the bound level deeper (Appendix~\ref{sm:wallchecks}).
The force on the wall grows when the instanton is placed closer to it, the opposite of a stress stored in the layers between, so this is an interaction of the instanton with the wall.
The wall that minimizes the bulk action is therefore not the wall the holonomy prescription gives, and it need not be, for two reasons.
The bulk action is not a term of the Hamiltonian, so the bulk's preference carries no energy.
And what the boundary sees of the wall passes through the Polyakov loops, which the prescription holds fixed, so the subtracted sheet difference is insensitive to the wall's rotation at the amplitude the bulk prefers (Appendix~\ref{sm:wallchecks}). The one quantity that moves with the wall is the bound level, a bulk quantity.

\section{Numerical methods}
\label{app:numerics}
This appendix defines the lattice and operators and contains every number in the paper. It describes the computations of the stationary bulk, the tests of the gauge fixing, of the Pauli-Villars subtraction and of the far wall, and the remaining results.

\subsection{Lattice and operators}
\label{sm:lattice}
The slabs used most often are $L_b^3\times L_s$ with $L_b=6$ and $L_s=12$ and $16$, with $L_s=8$ for the far-wall loops of Appendix~\ref{app:wall} and $8^3\times8$ and $10^3\times8$ for the larger windings and the two-winding paths of Table~\ref{tab:runs}. The smaller lattices used for individual checks are named where they appear.
The winding is $X(\lambda)$, $0\le\lambda\le1$, the tree-fixed rods of the links $\big(\Omega(x)\Omega(y)^\dagger\big)^\lambda$, with $\Omega(x)=\exp[i\pi(1-|x|/\rho)\,\hat x\cdot\sigma]$ for $|x|<\rho$ and $\Omega=1$ outside, $|x|$ the distance from the center of the boundary and $\rho$ the radius of the winding. $\Omega$ is a hedgehog of degree one, with $\Omega(0)=-1$, so at $\lambda=1$ the links are a large gauge transformation of the trivial ones and the rods return to the identity.
The lattice spacing is $a=1$ in all four directions, the boundary is periodic in the three transverse directions and the fermion antiperiodic in them.
The domain-wall mass is $M=1$ and the Wilson parameter $r=1$, and the bound level of the bulk instanton is scanned in $M$ from $0.5$ to $2$ on the same configurations.
The fermion is the five-dimensional Wilson operator in Hamiltonian form, in the gauge-fixed variables of Eq.~\eqref{eq:H},
\begin{align}
h=\sum_\ell\big(\chi^\dagger_x\Gamma_\ell U_\ell\chi_y+{\rm h.c.}\big)+\sum_x\chi^\dagger_x\gamma_0(M-4r)\chi_x\,,\qquad
\Gamma_\ell=\tfrac12\gamma_0\big(\gamma_d+r\big)
\label{eq:hwilson}
\end{align}
on the link from $x$ to $y=x+\hat d$, with
\begin{align}
\gamma_0=\mathrm{diag}(1,1,-1,-1)\,,\qquad
\gamma_{k}=\begin{pmatrix}0&-i\sigma_k\\ i\sigma_k&0\end{pmatrix}\ (k=1,2,3)\,,\qquad
\gamma_4=\begin{pmatrix}0&1\\1&0\end{pmatrix}
\end{align}
in the Dirac representation and $\gamma_4$ along $s$.
The links along $s$ are the identity and the mass is uniform, the walls being the two ends of the open direction, and $U_\ell$ is the color matrix on link $\ell$ of Eq.~\eqref{eq:U}.
The Pauli-Villars closure is the one of Appendix~\ref{sm:pvop}, antiperiodic in all four directions.
The magnetic term is $(1/g^2)\sum_\square(N-\mathrm{Re}\Tr P_\square)$ and the electric term $(g^2/2)\sum_\ell\mathcal J_\ell^2$, the Kogut-Susskind normalization, in which the couplings $g^2=4$, $8$ and $16$ of Appendix~\ref{sm:maxtreenum} are quoted.

\subsection{Spectra and sea energies}
\label{sm:spectra}
Single-particle spectra are dense diagonalizations of Eq.~\eqref{eq:hwilson} on the slab, $13\,824$ to $27\,648$ modes, and on the closed lattice for the partners. The levels near zero on $8^3$ and $10^3$ are from a shift-invert Lanczos solver.
The levels nearest zero along paths in rod space use shift-invert Lanczos and are followed from point to point by eigenvector overlap. Where both solvers were applied to one configuration they agree to $2\times10^{-12}$.
The filling of a sea is all negative levels, the Dirac vacuum of the sheet, with weight one half on the closed lattice.
For the state a winding reaches, the filling is the lowest $N_0$ levels, where $N_0$ is the number of negative levels on the trivial sheet.
Exact zero modes, which occur only with periodic transverse fermions, are filled to half their number.
Sea energies are sums of filled levels, and Berry phases are products of overlaps of the filled sea between successive points of a loop, closed on the first point, with the gap at the filling monitored along the loop.

\subsection{The bulk computation}
\label{sm:bulkcomp}
\paragraph{The flow.}
The bulk is the fixed point of the gradient flow Eq.~\eqref{eq:flow} on the links with an end in the bulk, with the boundary held fixed and the far wall frozen at the identity. The identity is within $0.08$ of the wall the prescription of Appendix~\ref{app:wall} assigns along the windings.
The integrator is third-order Runge-Kutta~\cite{Luscher:2010iy} with step $0.02$, halved whenever a step would carry a plaquette past the bound $\epsilon$ of the action below.
For a winding the boundary is brought to its target in small steps with a relaxation after each, so that the bulk follows one branch, and the flow is then run for $150$ to $600$ units.
The mid-slab configuration reached at $600$ units is a fixed point of the flow to $10^{-14}$ in the force, unchanged by a further $3000$ units, and a minimum, so the sea energies computed on it carry no flow or solver error at the digits quoted. Their differences are finite-$L_s$ and finite-$\epsilon$ effects.
Halving the flow step changes a relaxed action by $3\times10^{-9}$, its charge by $2\times10^{-12}$ and its links by $10^{-7}$ rms.

\paragraph{The instanton's position and charge.}
A wound sheet is started from the configuration reached by following a winding of the boundary, in which the lump sits at depth $3.6$ on $6^3\times12$, translated to mid-slab and relaxed again without a constraint.
The action changes by one percent between depths $3.6$ and $7.5$, and the instanton stays where it is placed, lattice pinning on a nearly flat direction, so the constrained stationary point of the definition agrees with it to that pinning.
Its position is the centroid of the layer-resolved clover charge.
The clover charge of the instanton is $0.7$ at this size, the usual deficit of the clover definition for a lump of a few spacings, while L\"uscher's geometric charge~\cite{Luscher:1981zq} of the same configuration, closed in $s$ with an identity seam, is $1.000000$ and the fermion index on the wound sheet is one.
The sector label was tested on random boundaries, smooth, rough and with net flux through a plane, six of each, and a sector-$n$ bulk was found on every one.
On one boundary of each kind, smooth, rough with $s_\square$ up to $0.32$, and with plane flux $1.6$ and $2.7$, the wound bulk binds one level per doublet at $\epsilon_0=-0.30$, $-0.34$, $-0.38$ and $-0.45$, with $99.8\%$ of its weight on layers $3$ to $8$ and below $5\times10^{-5}$ on the boundary and on the wall.

\paragraph{The action that holds the instanton.}
The flow must hold a lump of unit charge stationary. Which action does so is a choice of the classical computation, since the Hamiltonian contains no bulk plaquette term and the bulk enters it only through $F^{(n)}$ and the screened Coulomb term. The parameter $\epsilon$ below enters only that computation.
Every configuration in this paper was relaxed with L\"uscher's action~\cite{Luscher:1998du},
\begin{align}
S=\frac{1}{g^2}\sum_\square\frac{s_\square}{1-s_\square/\epsilon}\,,\qquad s_\square=1-\tfrac12\mathrm{Re}\Tr P_\square\,,
\label{eq:luscher_action}
\end{align}
at $\epsilon=0.4$, summed over the plaquettes that contain a bulk link, so the boundary's own plaquettes do not enter the flow. Its cost diverges as a plaquette approaches $\epsilon$, so it raises the price of the core plaquettes as a lump shrinks and holds the lump at a size set by $\epsilon$.
L\"uscher's theorem~\cite{Luscher:1981zq,Phillips:1986qd} needs $\|1-P_\square\|<0.173$ in the operator norm, $s_\square<0.015$ for \sutwo{}. Below that bound the geometric charge is constant along every path and the sector label is a theorem, and above it the exceptional set of the charge's definition, a wall of codimension one in the space of closed four-dimensional configurations, enters the admissible region.
At $\epsilon=0.4$ the configuration lies outside L\"uscher's bound, so its sector is protected by the divergence of the action at $s_\square=\epsilon$, which the flow cannot cross, and not by the theorem. It is nonetheless a local minimum, returning from random displacements of every bulk link by $0.1$, and a non-exceptional configuration with geometric charge exactly one.
The same configuration is reached without the winding. A seed carrying a unit of charge, layer $s$ holding the tree-fixed rods of $(\Omega(x)\Omega(y)^\dagger)^{\lambda(s)}$ with $\lambda$ falling from $1$ at the boundary to $0$ at the wall, relaxed with the trivial boundary and the flat wall, arrives at it to $4\times10^{-9}$ in the action, $6\times10^{-13}$ in every level near zero and $3\times10^{-13}$ in the charge per layer, for a linear and a cosine profile and for the seed's center shifted by one site.
At $\lambda=0.24$, $0.50$ and $0.74$ of the same winding, a seed was built from the boundary at that $\lambda$ and continued linearly to the identity at the wall. Relaxed at that boundary, it reaches the continued bulk to $10^{-8}$ in the action, $10^{-12}$ in the $72$ levels nearest zero and $10^{-13}$ in the charge per layer. The geometric charge of the slab closed in $s$ with an identity seam, which is $n$ plus the nearest integer to $\NCS$ of the boundary, is $0.000$, $0.500$ and $1.000$ at the three points. The middle value is at the sphaleron, where the rounding is undefined.
A contractible loop of the boundary, a winding and its inverse at two places traversed together on $8^3\times8$ and $10^3\times8$, returns the bulk to the trivial one with action and charge zero to $10^{-9}$, while the levels nearest zero approach each other to $0.06$ and separate again without crossing. Traversed one after the other on $10^3\times8$, the level bound to the first instanton returns to the band edge when the second winding undoes it.
A stable non-exceptional lump of integer charge, reached from a seed, that a contractible loop does not leave behind, is what supports the label at this $\epsilon$.

\paragraph{Choice of $\epsilon$.}
A winding inside L\"uscher's bound needs a lattice of linear size $\gtrsim14$.
A winding of radius $\rho$ has field strength of order $1/\rho^2$, and its largest boundary plaquette, at the sphaleron $\lambda=1/2$, falls off as $s_\square\approx11/\rho^4$. The windings used, $\rho=2.5$ on $6^3$ and $8^3$, $3$ on $8^3$ and $3.5$ on $10^3$, reach $0.25$, $0.14$ and $0.08$, with $s_\square\rho^4=9.8$, $11.0$ and $11.6$.
L\"uscher's bound then needs $\rho\approx5$, and the stricter bound of Ref.~\cite{Abe:2023ubg} $\rho\approx8$, and the bulk instanton, whose largest plaquette is $0.11$, would need a core about twice its present size.
An admissible magnetic term in the Hamiltonian would confine the rods to the bound at the price of $\beta=4/g^2\gg270$, and is not needed, since the register of Appendix~\ref{sm:register} keeps the sector unchanged across a boundary plaquette at $-1$.
Table~\ref{tab:runs} lists the admissibility data of the configurations used.

\paragraph{The Wilson action.}
The Wilson action has no stationary lump on these lattices. Flowed with it, the same lumps shrink from the first step with no plateau, the clover charge halving after $1.5$, $1.5$ and $2.0$ flow units on $6^3\times12$, $8^3\times8$ and $10^3\times8$ for the $\epsilon=0.4$ lumps and after $3.5$ and $4.5$ units for the larger lumps relaxed at $\epsilon=0.2$ and $0.12$ on $10^3\times8$.
The lifetime grows roughly as the cube of the lump's size, as a shrinking force of order $a^2/\rho^3$ predicts, and $g^2$ only rescales the flow time.
The Wilson flow with a fixed stopping time $\tau^*$ therefore needs a lump of $\rho\gtrsim5$ spacings, within reach of a $20^3$ boundary, at the price of a bulk that depends on $\tau^*$. The dense fermion diagonalization behind the sea energies, not the gauge flow, limited those to $6^3$, and the level computations to $10^3$.
\begin{table}[h]
\centering
\small
\begin{tabular}{l|cc|cc|ccl}
lattice & $\rho$ & $\epsilon$ & $s_\square^{\max}$ bdry & $s_\square^{\max}$ bulk & $Q_{\rm clover}$ & $\epsilon_0$ & sheet diff.\\
\hline
$6^3\times12$ & 2.5 & 0.4 & 0.25 & 0.11 & 0.68 & $-0.2715$ & $-2.7\times10^{-4}$\\
$6^3\times16$ & 2.5 & 0.4 & 0.25 & 0.11 & 0.69 & $-0.2865$ & $-1.3\times10^{-4}$\\
$8^3\times8$  & 2.5 & 0.4 & 0.25 & 0.10 & 0.71 & $-0.323$ & \\
$8^3\times8$  & 3   & 0.4 & 0.14 & 0.16 & 0 & & contractible pair\\
$10^3\times8$ & 3.5 & 0.4 & 0.08 & 0.09 & 0.73 & $-0.361$ & \\
$10^3\times8$ & 3.5 & 0.2 & 0.08 & 0.05 & 0.81 & $-0.514$ & \\
$10^3\times8$ & 3.5 & 0.12 & 0.08 & 0.03 & 0.84 & $-0.584$ & \\
$10^3\times8$ & 3.5 & 0.4 & 0.08 & & 0 & & sequential pair\\
\end{tabular}
\caption{Admissibility data of the computed configurations. The boundary maximum is at the sphaleron of the winding, the bulk maximum on the wound sheet after relaxation. $Q_{\rm clover}$ is the clover charge of the bulk instanton at mid-slab (its geometric charge is one), $\epsilon_0$ the bound level at $M=1$, and the sheet difference the Pauli-Villars-subtracted Dirac-vacuum energy of the wound sheet relative to the trivial one. On $10^3\times8$ the same winding was relaxed at three values of $\epsilon$. The flow stays inside the bound at each, the bound level deepens as the bound tightens, and at $\epsilon=0.12$ it lies below the boundary band edge of $0.535$. One level per doublet crosses on every winding listed, and none on the pairs. The contractible pair is a winding and its inverse at two places traversed together, the sequential pair the same two traversed one after the other (Appendix~\ref{sm:bulkcomp}).}
\label{tab:runs}
\end{table}

\subsection{Numerical tests of the gauge fixing and the bulk map}
\label{sm:maxtreenum}
\paragraph{Gauge-fixed against unfixed.}
The gauge-fixed Hamiltonian with fermions was compared with the unfixed Kogut-Susskind one, with Gauss's law imposed by a penalty, on one plaquette with a bulk layer and on $2\times1$ and $1\times2$ strips with comb trees and one doublet per site.
The full Gauss-law generator $G+Q$ of Eq.~\eqref{sm:gauss}, link part plus fermion charge, commutes with the Hamiltonian to $5\times10^{-15}$ on the strips and to $6\times10^{-12}$ on the plaquette, while the sign controls, $+Q_S$ in Eq.~\eqref{sm:J} or $L-Q$ in Eq.~\eqref{eq:global}, shift the lowest levels by $0.1$ to $0.6$.
The lowest levels agree to better than $10^{-5}$ on the plaquette at rod angular momentum $j\le1$, and on the strip to $1.1\times10^{-2}$ at $g^2=4$ and $1.4\times10^{-3}$ at $g^2=8$, the residual falling with the unfixed links' truncation at $j\le\tfrac12$. The two strips give the same spectrum to all printed digits.

\paragraph{The screened coupling.}
With the shift of Eq.~\eqref{sm:DX} in place, on one plaquette with one bulk layer, $[H,G+Q]$ vanishes to $9\times10^{-14}$ in the unfixed form and $[H,L_{\rm tot}+Q_{\rm tot}]$ to $7\times10^{-13}$ in the fixed one, independent of the tree, while a control without the adjoint transport in Eq.~\eqref{sm:DX} fails by $O(10^2)$.

\paragraph{Covariance and the orientation.}
Covariance, Eq.~\eqref{eq:equivariant}, was checked at a generic boundary on $6^3\times8$ by relaxing the bulk over the rods rotated by $h$. The link difference from the rotated bulk falls from $0.35$ to $5\times10^{-3}$ over $300$ flow units and is still falling, with the fitted rotation equal to $h$.
Along rays toward the trivial boundary the shape of the bulk follows the rods continuously while its orientation converges only for $\delta\gtrsim0.3$ and keeps drifting for $\delta\lesssim0.1$, as Appendix~\ref{sm:orientation} says.

\subsection{Numerical tests of the Pauli-Villars subtraction}
\label{sm:pvchecks}
\paragraph{The operator.}
On random backgrounds the spectrum of $h_{\rm PV}$ is symmetric about zero to $10^{-13}$.
Along the computed winding on $6^3\times12$ at $\epsilon=0.4$ the $\pm$ pair does not reach zero. Its splitting is $0.34$ at its narrowest, at the crossing, and follows the depth of the bound level, never below $0.53$ at $M=1.5$ and narrowing to $0.09$ at $M=0.75$.
The window in which the closure stays gapped along the winding is therefore $M\gtrsim1$, narrower than the index window, the range of $M$ in which the instanton binds a level, and this gap against the rod frequency is what the adiabatic treatment of the partners requires.
Ref.~\cite{Neuberger:1999pz} bounds the spectrum of the Wilson operator from below for all fields whose plaquettes are close enough to the identity, which would make the gap a theorem on every sheet and path. That bound needs $s_\square$ of order $10^{-3}$, far inside the admissibility bound used here, so at $\epsilon=0.4$ the gap is measured instead.

\paragraph{Backgrounds constant in $s$.}
On a background that does not depend on $s$ the Euclidean statements about the subtraction are statements about ground-state energies, and can be checked against Eq.~\eqref{eq:pv_static} directly.
Two were checked on $3^3\times L_s$ at $M=1$, the trivial background and a flat holonomy $\theta_x$ carried by the boundary, every bulk layer and the wall alike, whose dependence on $\theta_x$ is the potential of the walls for the Polyakov loop.
Each of $E_f$ and $E_{\rm PV}$ changes by $369$ per layer, while their difference converges exponentially in $L_s$ to an $L_s$-independent value, to five digits from $L_s=6$ on (Table~\ref{tab:pv_const}).
The subtracted potential $V(\theta_x)=E(\theta_x)-E(0)$ converges to $0.0375$ at $\theta_x=0.6$ and $0.1395$ at $\theta_x=1.2$, consistent with $0.10\,\theta_x^2$ and a small quartic, while $E_f(\theta_x)-E_f(0)$ alone grows with $L_s$. This is the finite potential of the walls that the domain-wall construction gives as $L_s\to\infty$.
\begin{table}[h]
\centering
\begin{tabular}{c|ccc|cc}
 & \multicolumn{3}{c|}{$E_f-E_{\rm PV}$} & \multicolumn{2}{c}{lowest $|w|$}\\
$L_s$ & $\theta_x=0$ & $\theta_x=0.6$ & $\theta_x=1.2$ & slab & closed\\
\hline
4  & 11.70122 & 11.73879 & 11.84088 & 1.692 & 1.652\\
6  & 11.70018 & 11.73771 & 11.83969 & 1.643 & 1.613\\
8  & 11.70017 & 11.73771 & 11.83969 & 1.621 & 1.599\\
12 & 11.70017 & 11.73771 & 11.83969 & 1.602 & 1.589
\end{tabular}
\caption{Pauli-Villars-subtracted ground-state energy $E_f-E_{\rm PV}$ on $3^3\times L_s$ at $M=1$ for backgrounds constant in $s$: trivial ($\theta_x=0$) and with a flat holonomy $\theta_x$ on every layer. Each of $E_f$ and $E_{\rm PV}$ is of order $-370\,L_s$. The last two columns are the smallest single-particle level of the slab and of the closed lattice of length $2L_s$ at $\theta_x=0$.}
\label{tab:pv_const}
\end{table}

\paragraph{The bulk instanton.}
With the bulk instanton at mid-slab, on $6^3\times12$ and $6^3\times16$ at $M=1$, the closed lattice binds the $\pm$ pair at $\mp0.2715$ and $\mp0.2865$, the slab's bound level is at $-0.2715$ and $-0.2865$ with the same profile in $s$, and the rest of the closed lattice's spectrum stays gapped.
Comparing Dirac vacua, the sea part of the sheet difference is $1.43$ and $1.45$ before subtraction and $-3\times10^{-4}$ and $-1\times10^{-4}$ after it, and on the configurations with the instanton at depth $3.6$ it is $-1\times10^{-3}$, $-6\times10^{-4}$ and $-2\times10^{-3}$ at $M=0.5$, $1$ and $2$.

\paragraph{The state at fixed fermion number.}
The Dirac vacuum of the wound sheet has one fermion more than that of the trivial sheet, so at fixed fermion number the wound state has a hole in its highest filled level.
With antiperiodic transverse fermions on $6^3$ the boundary's lowest level is $0.87$, above the bound level, so the hole is on the bound level and the state lies $0.271$ and $0.286$ above the trivial sheet after subtraction, equal to $|\epsilon_0|$ to $10^{-3}$.
With periodic transverse fermions the boundary and the mirror each have four exact zero modes, the bound level is at $-0.301$, the hole is in a boundary zero mode, and the cost is zero to $10^{-5}$.
With $L_b\gtrsim20$ the antiperiodic boundary also has a level below $|\epsilon_0|$.

\paragraph{The bound level.}
On the mid-slab configuration $\epsilon_0=+0.073$, $-0.112$, $-0.271$, $-0.434$ and $-0.169$ at $M=0.5$, $0.75$, $1$, $1.5$ and $2$, where at $M=2$ the bulk gap has closed to $0.25$.
It crosses zero at the lower edge of the index window and deepens with $M$ through $M=1.5$, as the effective Wilson mass of the instanton's zero mode, $-M$ plus a Wilson shift of order $1/\rho_I^2$. A linear fit over $0.5\le M\le1.5$ gives $\epsilon_0\approx-0.50M+0.28$ ($\rho_I\approx1.9$), and over $0.75\le M\le1.5$ $-0.42M+0.18$ ($\rho_I\approx2.4$), with a slope well below the first-order $-1$ in either window.
The level's weight on the far wall is $1.3\times10^{-4}$, $1.1\times10^{-5}$, $1.9\times10^{-7}$ and $7.7\times10^{-11}$ at lump-to-wall distances of $4$, $5$, $8$ and $11$ layers, on $L_s=8$, $12$, $16$ and a $6^3\times20$ slab made by padding the $L_s=16$ one with four flat layers.
A fit to $e^{-2\kappa d}$ gives $\kappa=1.0$, against $1.0$ to $1.15$ read from the profiles, which fall by a factor of ten per layer on both sides of the core, and the level's energy at those four distances is $-0.229$, $-0.272$, $-0.287$ and $-0.292$.

\paragraph{The electric part of the sheet difference.}
The Coulomb energy of the sea's charge correlations is $\sum_{xy}G_{xy}\langle Q_x\cdot Q_y\rangle$, with $G_{xy}$ the screened Green's function of Eq.~\eqref{eq:Phi} and the zero mode removed.
Its difference between the sheets, with the Pauli-Villars counterpart subtracted, is $-2\times10^{-5}$ and $-1\times10^{-5}$ on $6^3\times12$ and $6^3\times16$ with the screened coupling of the main text, the same for the heat-kernel, Gaussian and elliptic profiles, and $-0.021$ and $-0.022$ with the point coupling $f_s=1$.

\subsection{Numerical tests of the far wall}
\label{sm:wallchecks}
\paragraph{The prescription on the windings.}
On the computed windings on $6^3\times12$ the averaged prescription gives a wall within $0.08$ of the identity at every $\lambda$, and the identity at the ends. The winding's lines carry rotations about all axes, and their vectors cancel in the average. The loops through a single site, by contrast, reach $|\phi|=1.44$ at the sphaleron.
The computations used the wall at the identity. The crossing is the same with the averaged prescription's wall, the level nearest zero being $-0.0173$ against $-0.0178$ at $\lambda=0.74$ and $0.2210$ against $0.221$ at $\lambda=0.50$.

\paragraph{The mismatch.}
With the wound bulk at mid-slab and the wall turned about the instanton's axis at fixed boundary, the action falls by $0.45$ and $0.6$ per radian at $L_s=12$ and $16$.
With the wall's three holonomies free to move under the flow, each along its own axis by the mean force on it, the wall settles $1.5$ to $2$ radians away, the action is lower by $0.6$ to $0.9$, a few percent of the instanton's action, the bound level deepens from $-0.27$ to $-0.44$, and the instanton moves by about half a layer with its preference for a depth intact. The relaxed wall's dependence on $L_b$ has not been measured.
With the wall turned by $0.1$ and $0.2$ radians the subtracted Dirac-vacuum sheet difference changes by $4\times10^{-5}$ at $L_s=12$ and $16$.
Turned by $1.5$ radians, the amplitude the bulk prefers, it changes by $3.5\times10^{-4}$ at $L_s=12$ and $3.0\times10^{-4}$ at $L_s=16$, while the fixed-fermion-number difference, which exceeds the Dirac-vacuum difference by $|\epsilon_0|$ on these lattices, moves from $0.271$ to $0.394$ and from $0.286$ to $0.426$ with the bound level.
On $8^3\times8$ the turn by $1.5$ radians changes the Dirac-vacuum difference by $2.4\times10^{-4}$ and deepens the bound level from $-0.32$ to $-0.41$, so the boundary's insensitivity does not weaken with $L_b$.

\paragraph{The singular set.}
Berry phases of the filled sea were computed on $6^3\times8$ for five loops in rod space that encircle the singular set and for two control loops that encircle nothing.
The boundary is a field of constant links whose three Polyakov loops circle the singular point, the wall is given by the prescription, and the bulk is interpolated along $s$ between the two on a fixed random background, so this tests the prescription's singular set and not the stationary bulk.
Each loop has $96$ points. The largest link step is $0.10$ and $0.09$ on two of the encircling loops and $0.03$ and $0.04$ on the controls. On the other three encircling loops it is $0.13$ to $0.16$, above the step the computation was set up to trust, and their readings carry that caveat.
The phase of the filled sea is zero on the controls, with gaps of $1.30$ and $0.76$, and on four of the encircling loops, with gaps of $1.24$, $0.91$, $0.76$ and $0.04$.
On the fifth, at a boundary with large non-commuting flux, one level with $99\%$ of its weight on the boundary layer passes through zero and back along the loop, so the ground state changes its filling on a segment of the loop.
The phases at the two fixed fillings on either side of that level differ by $\pi$, carried by that level alone, with gaps of $0.08$ and $0.01$ at those fillings.
The $\pi$ is the same whether the wall is tied to the rods, held at its first value, or set to the identity, and the crossing is the same with the wall held at its first value, so both belong to the boundary and not to the prescription.
The Pauli-Villars sea, whose connection is separate (Appendix~\ref{app:pv}), was not included in these loops.

\subsection{Results}
\label{sm:results}
\paragraph{The $n=2$ sheet.}
A second winding over the $n=1$ bulk on $6^3\times12$ gives two units side by side at $s=5$--$6$, with action $40.63$, three percent below twice the single unit's $20.97$, and two bound levels at $-0.405$, one per unit.
After the Pauli-Villars subtraction its Dirac vacuum is degenerate with the trivial one to $1.4\times10^{-3}$, and the state at fixed fermion number lies $0.81$ above it, twice the $n=2$ bound level.

\paragraph{Occupations.}
Where the missing fermions sit on a wound sheet depends on the volume.
Table~\ref{tab:occ} lists the occupations and the center parity $(-1)^F$ of the boundary for the states discussed in the main text.
\begin{table}[h]
\centering
\begin{tabular}{lccc}
\hline
state ($N_f=1$) & boundary & bound level & parity\\
\hline
trivial Dirac vacuum & $N_0^{\rm bdy}$ & -- & even\\
wound Dirac vacuum, $N+1$ fermions & $N_0^{\rm bdy}$ & filled & even\\
wound, fixed $N$, small box & $N_0^{\rm bdy}$ & empty & even\\
wound, fixed $N$, large box & $N_0^{\rm bdy}-1$ & filled & odd\\
\hline
\end{tabular}
\caption{Occupations on the two sheets ($6^3\times12$, $M=1$). The mirror is in its sea in every row. $N_0^{\rm bdy}$ is the number of boundary levels filled on the trivial sheet. The small box is one whose lowest boundary level lies above $|\epsilon_0|$, so the hole sits on the bound level. In the large box it sits on the boundary. The parity is $(-1)^F$ of the boundary relative to the trivial sheet, and the boundary count is sharp only away from the crossing, where the level's weight moves from the boundary to the core (Fig.~\ref{fig:geom}b).}
\label{tab:occ}
\end{table}

\paragraph{The perturbative criterion in a $U(1)$ slab.}
The locality criterion of Refs.~\cite{Grabowska:2015qpk,Grabowska:2016bis} was tested in the Hamiltonian with no dynamics.
A winding of the boundary field moves fermion number into the bulk species by species, and the question is whether the gauge charge that arrives there vanishes, since by Gauss's law that is the only part the boundary can see.
The lattice is a ring of $L_x=24$ sites times $s=0,\dots,L_s-1$ with $L_s=12$ and $20$, with two-dimensional Wilson fermions at $(M,r)=(1,1)$, antiperiodic on the ring.
A species of charge $q$ couples through $U^q$, and the chirality of its boundary mode is set by the sign of $(M,r)$.
The boundary layer is fixed to a maximal tree of the ring, so its one physical variable is the holonomy $\theta$ on the wrap link, and the winding is $\theta=2\pi\lambda$ in $100$ steps.
The bulk is an earlier form of the construction than the stationary bulk of the main text, a copy of the boundary layer for $s<4$, a middle for $4\le s\le7$ ($4\le s\le11$ at $L_s=20$) whose links are relaxed, and a far region at the identity, and the middle is built in two ways that differ exactly in whether the bulk keeps the flux quantum from one step to the next.
In the first, the continued form, it is relaxed from the previous step with the action of Eq.~\eqref{eq:luscher_action} at $\epsilon=0.6$, so that the flux quantum pushed in by the winding stays.
In the second, the fresh form, it is relaxed from zero at every $\lambda$ with the Wilson action, a single-valued function of $\theta$ whose flux jumps from $+\tfrac12$ to $-\tfrac12$ at $\lambda=0.49$ and returns to zero.
The contents are one Weyl fermion of charge one, which is anomalous, and the $3$-$4$-$5$-$0$ set, charges $3$ and $4$ of one chirality and $5$ and $0$ of the other, which is not. The charge-zero fermion balances the number of left and right movers and plays no part here.
At each $\lambda$ the ground state is the filled negative sea of every species, and the gauge charge $\sum_iq_i\,\delta N_i$ of the bulk relative to $\lambda=0$ is read off layer by layer (Table~\ref{tab:u1slab}).
In the continued form each species moves $q$ units of its own number into the bulk per winding, with the same profile in $s$ for every species.
For the anomaly-free content the gauge charge of that transfer is zero at every point of the winding in both forms, $3\times10^{-4}$ at the end from the exponential tails, so the boundary cannot tell from its own Gauss's law how the middle was built.
For the single Weyl fermion the bulk carries gauge charge $\lambda$ in the continued form and ends with one unit stored with the flux quantum, $0.79$ of it in the middle at $L_s=12$ and $0.87$ at $L_s=20$, while in the fresh form the bulk charge rises to $+\tfrac12$, flips to $-\tfrac12$ where the middle jumps, and returns to zero.
The charge that the boundary's own Gauss law assigns to the bulk depends on how the bulk is built exactly when the anomaly does not cancel, and at every $\lambda$. 
\begin{table}[h]
\centering
\begin{tabular}{ll|ccc}
\hline
content & middle & $Q_{\rm bulk}(\tfrac12)$ & $Q_{\rm bulk}(1)$ & flux$(1)$\\
\hline
anomalous, one Weyl $q=1$ & continued & $+0.500$  & $+0.999$  & $1$\\
anomalous, one Weyl $q=1$ & fresh     & $-0.500$  & $0$       & $0$\\
anomaly free, $3$-$4$-$5$-$0$ & continued & $+0.0001$ & $+0.0003$ & $1$\\
anomaly free, $3$-$4$-$5$-$0$ & fresh     & $0$       & $0$       & $0$\\
\hline
\end{tabular}
\caption{The $U(1)$ slab with a $1+1$-dimensional boundary, $L_x=24$, $L_s=12$ ($L_s=20$ agrees to the digits shown). Columns: the gauge charge of the bulk relative to $\lambda=0$ at the middle and at the end of the winding, and the flux quantum held by the middle at the end. In the continued form each species moves $q$ units of its own number into the bulk, $(3,4,-5,0)$ for the anomaly-free set and $1$ for the single fermion. In the fresh form none does. The gauge charge of that transfer, $9+16-25+0$, is zero for the anomaly-free set in both forms, so only the single fermion's bulk charge depends on the form.}
\label{tab:u1slab}
\end{table}

\paragraph{The holonomy of the global constraint.}
The loop is a $2\pi$ rotation of the rods about a fixed axis, $X_\kappa\to R(t)X_\kappa R(t)^{-1}$, a closed loop that is non-contractible as a loop in $SO(3)$ and contractible in rod space, with the bulk following by covariance.
On $6^3\times8$ at a generic boundary the Berry phase of the Dirac vacuum is $0$ on the trivial sheet and $\pi$ on the wound sheet, where one more level is filled. It is $N\pi$ at every filling $N$ on both sheets. The spectrum returns to $10^{-14}$ and the overlaps between successive points stay above $0.75$.
The relative holonomy of the two sheets is $(-1)^{N_f}$ for $N_f=1$, the sign of Witten's condition. A random $s$-dependent background on $3^3\times6$ gives the same parity rule.

\paragraph{The inertia correction.}
$\mathcal I^{\rm PV}$ of Appendix~\ref{sm:pvdyn} was computed as the full matrix over every boundary link direction for one color generator on $3^3$, $4^3$ and $5^3$ boundaries for $L_s$ from $4$ to $16$, with the bulk map linearized about the trivial sheet, restricted to the rods of the comb tree and diagonalized together with the rod inertia $\mathcal G$ of Eq.~\eqref{eq:Ginertia}.
Its lowest eigenvalue is the Polyakov-loop mode, $-0.0063\,L_s$, $-0.0098\,L_s$ and $-0.0117\,L_s$ per rod on the three volumes, linear in $L_s$. With the transverse part alone $K>0$ holds for $g^2L_s\lesssim10^2$, and with the Coulomb part included it fails at $g^2L_s\approx5$ on $3^3$ and earlier on larger boundaries.
The inertia of both seas, $\mathcal I^f+\mathcal I^{\rm PV}$, has lowest eigenvalue $-0.0025$, $-0.0021$ and $-0.0009$ on the three volumes at every $L_s$, and the static response of the two seas together is unchanged to five digits between $L_s=4$ and $16$ and is $1$ to $8$ percent of either sea's alone, while each alone grows linearly with $L_s$.
In the gauge-fixed metric of Eq.~\eqref{eq:Ginertia}, with the Coulomb modes included, $\mathcal G+\mathcal I^{\rm PV}>0$ holds for $g^2<1.3$, $0.7$ and $0.5$ at $L_s=4$, $8$ and $12$ on $3^3$ and $g^2<0.5$ and $0.3$ at $L_s=4$ and $6$ on $4^3$, while $\mathcal G+\mathcal I^f+\mathcal I^{\rm PV}>0$ holds for $g^2<33$, $32$ and $31$ on $3^3$ and $g^2<12$ and $11$ on $4^3$.
With the layer closure, a circle of $2L_s-2$ layers with layers $0$ and $L_s-1$ appearing once, the lowest eigenvalue of $\mathcal I^f+\mathcal I^{\rm PV}$ is $+0.0008$ at $L_s=4$ and $8$ on $3^3$ and $+0.0012$ at $L_s=4$ on $4^3$. So $K>0$ at every $g^2$. The partners alone, $\mathcal G+\mathcal I^{\rm PV}>0$, hold only for $g^2<2.0$ and $0.9$ on $3^3$ and $0.7$ on $4^3$. 
With that closure $E_f-E_{\rm PV}$ on $3^3$ is $-357.153$, $L_s$-independent to five digits from $L_s=8$ on, as in Table~\ref{tab:pv_const}. It lies below the link closure's value by one layer's $369$. Its dependence on the holonomy $\theta_x$ is $0.18$ and $0.66$ at $\theta_x=0.6$ and $1.2$, against $0.04$ and $0.14$ for the link closure, the unsubtracted layer's own potential.

\end{document}